\documentclass[11pt,a4paper]{article}
\usepackage{a4wide}
\usepackage{lineno,hyperref}
\usepackage{booktabs}
\usepackage{amsmath}
\usepackage{amssymb}
\usepackage{bbold}
\usepackage{wasysym}
\usepackage[flushleft]{threeparttable}
\usepackage{lscape}
\usepackage{rotating}
\usepackage{upgreek}
\usepackage[english]{babel}
\usepackage{cite} 
\usepackage{multirow}
\usepackage{wrapfig}
\usepackage[section]{placeins}

\newcommand{\norm}[1]{\lVert#1\rVert}

\def\rms{{\rm s}}

\def\sl{\rms_\lambda}

\def\bsx{{\boldsymbol x}}
\def\bsxi{{\boldsymbol \xi}}

\def\proof{\noindent{\sl Proof:}\kern0.6em}

\def\dual{\mathstrut^*\kern-0.1em}

\def\lvec#1{\setbox0=\hbox{$#1$}
    \setbox1=\hbox{$\scriptstyle\leftarrow$}
    #1\kern-\wd0\smash{
    \raise\ht0\hbox{$\raise1pt\hbox{$\scriptstyle\leftarrow$}$}}
    \kern-\wd1\kern\wd0}
\def\rvec#1{\setbox0=\hbox{$#1$}
    \setbox1=\hbox{$\scriptstyle\rightarrow$}
    #1\kern-\wd0\smash{
    \raise\ht0\hbox{$\raise1pt\hbox{$\scriptstyle\rightarrow$}$}}
    \kern-\wd1\kern\wd0}
\def\slash#1{\setbox0=\hbox{$#1$}\setbox1=\hbox{$\kern1pt/$}
    #1\kern-\wd0\kern1pt/\kern-\wd1\kern\wd0}

\def\nab#1{{\nabla_{#1}}}
\def\nabstar#1{{\nabla\kern0.5pt\smash{\raise 4.5pt\hbox{$\ast$}}
               \kern-5.5pt_{#1}}}

\def\nabbarstar#1{{\overleftarrow{\nabla}\kern0.5pt\smash{\raise 4.5pt\hbox{$\ast$}}
               \kern-5.5pt_{#1}}}

\def\nabdbarstar#1{{\overleftrightarrow{\nabla}\kern0.5pt\smash{\raise 4.5pt\hbox{$\ast$}}
               \kern-5.5pt_{#1}}}

\def\drvstar#1{{\partial\kern0.5pt\smash{\raise 4.5pt\hbox{$\ast$}}
               \kern-6.0pt_{#1}}}

\def\ldrvstar#1{{\lvec{\,\partial}\kern-0.5pt\smash{\raise 4.5pt\hbox{$\ast$}}
               \kern-5.0pt_{#1}}}

\def\MSbar{\overline{\rm MS\kern-0.5pt}\kern0.5pt}

\def\psibar{\overline{\psi}}

\def\zetabar{\bar{\zeta}}
\def\zetaprime{\zeta\kern1pt'}
\def\zetabarprime{\zetabar\kern1pt'}

\def\dirac#1{\gamma_{#1}}
\def\diracstar#1#2{
    \setbox0=\hbox{$\gamma$}\setbox1=\hbox{$\gamma_{#1}$}
    \gamma_{#1}\kern-\wd1\kern\wd0
    \smash{\raise4.5pt\hbox{$\scriptstyle#2$}}}

\def\Tr{{\rm Tr}}

\def\Obs{{\mathcal O}}

\def\Ds{D_{\rm s}}
\def\DsdagDs{\Ds{\Ds}^{\kern-1pt\dagger}}

\def\avg#1{{\kern1.0pt\overline{\kern-1.0pt#1\kern-1.0pt}\kern1.0pt}}

\newcommand{\be}{\begin{equation}}
\newcommand{\ee}{\end{equation}}
\newcommand{\bea}{\begin{eqnarray}}
\newcommand{\eea}{\end{eqnarray}}

\newcommand{\E}{\text{e}}
\newcommand{\id}{1\!\!1}

\newcommand{\msbar}{{\rm \overline{MS\kern-0.05em}\kern0.05em}}

\newcommand{\ba}{\begin{eqnarray}}
\newcommand{\ea}{\end{eqnarray}}

\renewcommand{\vec}[1]{\boldsymbol{#1}}

\newcommand{\mcr}[1][\vphantom{()}]{m_{\rm cr}^{#1}}

\begin{document}
\begin{titlepage}
\begin{flushright}
\hfill CERN-TH-2021-211\\
\end{flushright}

\begin{center}
    
$\;\;\;$
\vspace{1.5cm}

{\Large\bf Non-perturbative thermal QCD at all temperatures:\\[0.25cm]
  the case of mesonic screening masses}

\end{center}
\vskip 0.75 cm
\begin{center}
{\large 
Mattia Dalla Brida$^{\scriptscriptstyle a}$, 
Leonardo Giusti$^{\scriptscriptstyle b,c}$, 
Tim Harris$^{\scriptscriptstyle d}$, \\[0.125cm]
Davide Laudicina$^{\scriptscriptstyle b,c}$, and
Michele Pepe$^{\scriptscriptstyle c}$}
\vskip 1.0cm
$^{\scriptstyle a}$ Theoretical Physics Department, CERN, \\
                CH-1211 Geneva 23, Switzerland\\
\vskip 0.25cm
$^{\scriptstyle b}$ Dipartimento di Fisica, Universit\`a di Milano-Bicocca,\\
                 Piazza della Scienza 3, I-20126 Milano, Italy\\ 
\vskip 0.25cm
$^{\scriptstyle c}$ INFN, Sezione di Milano-Bicocca,\\
                 Piazza della Scienza 3, I-20126 Milano, Italy\\
\vskip 0.25cm
$^{\scriptstyle d}$ School of Physics and Astronomy, University of Edinburgh, \\
                Edinburgh EH9 3JZ, UK
\vskip 0.5cm
{\bf Abstract}
\vskip 1.5ex
\end{center}

\noindent
We present a strategy based on the step-scaling technique to study non-perturbatively thermal QCD up to very
high temperatures. As a first concrete application, we compute the flavour non-singlet
meson screening masses at 12 temperatures covering the range from $T \sim 1$~GeV up
to $\sim 160$~GeV in the theory with three massless quarks. The calculation is carried
out by Monte Carlo simulations on the lattice by considering large spatial
extensions in order to have negligible finite volume effects. For each temperature we
have simulated 3 or 4 values of the lattice spacing, so as to perform the continuum
limit extrapolation with confidence at a few permille accuracy. Chiral symmetry restoration
manifests itself in our results through the degeneracy of the vector and the axial vector channels
and of the scalar and the pseudoscalar ones. In the entire range of temperatures explored,
the meson screening masses deviate from the free theory result, $2 \pi T$, by at most a
few percent. These deviations, however, cannot be explained by the known leading term
in the QCD coupling constant $g$ up to the highest temperature, where other contributions are
still very relevant. In particular the vector-pseudoscalar mass splitting turns out to be of
$O(g^4)$ in the entire range explored, and it remains clearly visible up to the highest
temperature, where the two screening masses are still significantly
different within our numerical precision. The pattern of different contributions that
we have found explains why it has been difficult in the past to match non-perturbative lattice
results at $T \sim 1$~GeV with the analytic behaviour at asymptotically high temperatures.

\vfill

\end{titlepage}


\section{Introduction}
Thermal Quantum Chromodynamics (QCD) plays a fundamental r\^ole in particle and nuclear
physics, and in cosmology. Apart from its intrinsic theoretical interest, the collective
behaviour of strongly-interacting particles is crucial input for determining the evolution
of the Universe in its early stages. Today the quark-gluon plasma is also produced and
investigated at heavy-ion colliders, where some of its basic properties are essential
to analyze the experimental data.

At asymptotically high temperatures, thermal QCD is described by a three-dimensional
effective gauge theory~\cite{Ginsparg:1980ef,Appelquist:1981vg} which needs to be solved
non-perturbatively~\cite{Linde:1980ts}. As a result, perturbation theory can predict the coefficients
of the expansion in the strong coupling constant $g$ only up to a finite order. An important
example is the Equation of State (EoS), where non-perturbative contributions start at $O(g^6)$
\cite{Braaten:1995jr,PhysRevD.67.105008}. In the SU(3) Yang--Mills theory, these terms
are found to be large up to temperatures two orders of magnitude higher than the
critical one~\cite{Giusti:2016iqr}. All these facts call for a non-perturbative strategy to
study thermal QCD up to very high temperatures, possibly up to the electroweak scale.

The purpose of this paper is to combine information encoded in the three-dimensional
effective theory, lattice QCD, step-scaling techniques, and Monte Carlo integration to devise
a strategy for studying QCD non-perturbatively up to very high temperatures from first
principles. The aim is to generalize to QCD the proposal made in Ref.~\cite{Giusti:2016iqr} for
the Yang--Mills theory. As a concrete implementation, we consider QCD with $N_f=3$ flavours
of massless quarks in the range of temperatures from about 1 GeV up to approximately 160
GeV\footnote{As the temperature becomes higher and higher, the relevance of the light quark masses
becomes quickly negligible.}.

As a first application we compute the non-singlet meson screening masses, maybe the
simplest properties of the plasma to be computed. They characterize the exponential
decay of two-point correlation functions of fermion bilinears in the spatial directions,
i.e. their inverses are the
long-distance spatial correlation lengths when mesons are present in the plasma.
Screening masses can be easily investigated numerically, they are related to spectral
functions, and they signal the restoration of chiral symmetry at high temperature.
Their $O(g^2)$ component is known, and is found
to be spin independent~\cite{Laine:2003bd}. The first spin-dependent term is expected to appear
only at $O(g^4)$ \cite{Koch:1992nx,Hansson:1991kb}. These masses are therefore ideal quantities to
test the strategy proposed in this paper, and to further investigate
the relevance of non-perturbative contributions in thermal QCD. It must be said that they 
have been computed non-perturbatively in lattice QCD for decades, see
Ref.~\cite{Cheng:2010fe,Brandt:2014uda,Bazavov:2019www,Brandt:2019ksy} for
recent efforts. These computations, however, are limited to temperatures up to approximately
$1$~GeV. Here we want to extend the range up to much higher temperatures so as to elucidate the
approach to the infinite temperature limit, where the effective theory is expected to match
thermal QCD.

The paper is organized as follows. In section~\ref{sec:stgy1} we review the three-dimensional
effective theory, and summarize the properties relevant to this paper. The next section
is devoted to introduce the screening masses, and to summarize the known analytic contribution
to them. In section~\ref{sec:lattice} we present the strategy for simulating lattice QCD up to very high
temperatures. The lattice setup and the raw results are discussed in section~\ref{sec:corfcn},
while the value of the screening masses in the continuum limit are given in section~\ref{sect:num_study}.
We discuss and interpret the final results in section~\ref{sec:disc}, while our conclusions and outlook
are given in section~\ref{sec:conc}. Notations, conventions, and technical details are reported in
several appendices.

\section{Preliminaries on the effective theory at large $T$ \label{sec:stgy1}}
In thermal QCD at high temperature the physics that takes place at energies much lower than $T$,
or equivalently that involve distances much larger than the temporal direction, can be described by
a three-dimensional effective gauge theory~\cite{Ginsparg:1980ef,Appelquist:1981vg,Braaten:1995jr}, dubbed
Electrostatic QCD (EQCD), defined by the action
\be
S_{\rm EQCD} = \frac{1}{g^2_{_{\rm E}}} \int d^3 x \Big\{\frac{1}{2} \Tr\left[F_{ij} F_{ij}\right] +
\Tr \left[(D_j A_0) (D_j A_0)\right] + m^2_{_{\rm E}} \Tr\left[A_0^2\right]\Big\} +\dots
\ee
where the dots stand for higher dimensional operators, see Ref.~\cite{Laine:2016hma} for
a recent review. The field content is made of the Matsubara zero-modes of the gauge field,
while quark fields are decoupled because their modes
pick up a mass proportional to $\pi T$ due to the antiperiodic boundary conditions in the
compact direction. The dynamics of the spatial components is governed by a three-dimensional
Yang-Mills theory with field strength tensor $F_{ij}$ and dimensionful coupling constant
$g^2_{_{\rm E}}$. The temporal component of the gauge field $A_0$ behaves as a three-dimensional
scalar field of mass $m_{_{\rm E}}$ which transforms under the adjoint representation of the
gauge group. The matching with QCD fixes the low-energy constants to be
$m^2_{_{\rm E}} = \frac{3}{2} g^2 T^2 + \dots$ and $g^2_{_{\rm E}} = g^2 T+\dots $, with $g$ being the renormalized
coupling of QCD (usually taken at scale $2\pi T$), and the dots stand for
higher order terms in the coupling constant~\cite{Laine:2005ai}.

At asymptotically high $T$, the coupling $g$ is small and three different energy
scales develop so that 
\be
\frac{g^2_{_{\rm E}}}{\pi} \ll m_{_{\rm E}} \ll \pi T\; .  
\ee
If one is interested in processes at scales of $O(g^2_{_{\rm E}})$, the scalar field can be integrated out.
The action of the remaining effective theory, dubbed Magnetostatic QCD (MQCD), is given by
\be
S_{\rm MQCD} = \frac{1}{g^2_{_{\rm E}}} \int d^3 x \Big\{\frac{1}{2} \Tr\left[F_{ij} F_{ij}\right] \Big\} +\dots
\ee
Being a three-dimensional Yang--Mills theory, it has non-perturbative dynamics
and therefore it needs to be solved non-perturbatively~\cite{Linde:1980ts}. All
dimensionful quantities are proportional to the appropriate power of $g^2_{_{\rm E}}$ times a
non-perturbative coefficient.

This in turn implies that, at asymptotically high temperatures, the mass gap developed by
thermal QCD is proportional to $g^2_{_{\rm E}}$~\cite{Arnold:1995bh}. Finite volume
effects are therefore expected to be exponentially small in $g^2_{_{\rm E}} L = g^2 T L+\dots$
times a non-perturbative coefficient, see below and appendix~\ref{app:finiteV}. This fact turns out to be crucial for the strategy
outlined in the next sections for studying non-perturbatively thermal QCD up to very high
temperatures. At intermediate temperatures, the pre-factors may be relevant in determining what is the
mass gap of the theory. But given the relevant scales in the problem, and provided
the temperature is sufficiently high with respect to $\Lambda_{\rm QCD}$, the mass gap of the theory is
always expected to be proportional to the temperature times an appropriate power of the coupling
constant~\cite{Laine:2009dh}.

\subsection{Fermion correlators}
In the effective field theory approach, the quarks are very heavy fields that can be considered, in first approximation, as
static fields. By adopting the notation of Ref.~\cite{Laine:2003bd}, we represent the spinor field as
\be
\psi=\left(\begin{array}{c}
\chi\\[0.25cm]
\phi
\end{array}\right)\, ,
\ee
so that the effective action for the fermion modes can be written
as~\cite{Hansson:1991kb,Koch:1992nx,Huang:1995tz,Laine:2003bd} 
\ba\label{eq:Seffq}
S^{\rm eff}_q = \int d^3 x \Big\{\hspace{-0.625cm}&& i \chi^\dagger \Big[M - g_{_{\rm E}} A_0 + D_{3} -
\frac{1}{2 M}\left(D_k^2 + \frac{g_{_{\rm E}}}{4 i}[\sigma_k,\sigma_l] F_{kl}\right) \Big]\chi +\\ \nonumber
&& i \phi^\dagger \Big[M - g_{_{\rm E}} A_0 - D_{3} -
\frac{1}{2 M}\left(D_k^2 + \frac{g_{_{\rm E}}}{4 i}[\sigma_k,\sigma_l] F_{kl}\right) \Big]\phi \Big\} + \dots
\ea
where the mass $M$ is identified with a low-energy constant which,
for the lightest modes, is given by $\displaystyle M=\pi T[1 + g^2/(6\pi^2)] +\dots $, while the spatial direction
$3$ is the one along which the mesonic 2-point functions are measured in order to compute the screening masses.
In three dimensions the chiral symmetry group is
enlarged with respect to the four-dimensional one due to dimensional reduction. The quark mass $M$, however,
breaks the three-dimensional chiral group down to the unbroken four-dimensional chiral symmetry, and the
familiar pattern of chiral symmetry restoration at high temperature is recovered~\cite{Appelquist:1986fd}.
As we shall see, the restoration of chiral symmetry in thermal QCD will show up in the degeneracy of
various masses measured by Monte Carlo simulations at large $T$.

By scrutinizing the magnitude of the various terms in Eq.~(\ref{eq:Seffq}), the interaction term in the
covariant derivatives $D_k$ ($k=1,2$) and the spin-dependent contributions proportional to $F_{kl}$ are of
higher order in the strong coupling constant with respect to the other terms, and they can be dropped
if one is interested in the leading contributions~\cite{Laine:2003bd}. Before doing so, however, it is
interesting to notice that the spin-dependent terms give contributions starting at $O(g^4)$~\cite{Koch:1992nx,Hansson:1991kb}.

\section{Definition of the mesonic screening masses \label{sec:scrM}}
We are interested in the screening masses related to flavour non-singlet fermion bilinear operators
\be\label{eq:bils}
{\cal O}^a(x) = \psibar(x) \Gamma_{{\cal O}} \, T^a \,\psi(x)\;,\\
\ee
where $\Gamma_{{\cal O}}=\left\{\id,\gamma_5,\gamma_\mu,\gamma_\mu\gamma_5\right\}$ characterizes the structure of
the operators in the Dirac space, with the latter named as usual as ${\cal O}=\left\{S,P,V_\mu,A_\mu\right\}$,
and we restrict ourselves to $\mu=2$.
Since we will be considering QCD with three massless flavours, the Hermitean matrices $T^a$ are the traceless generators of $SU(3)$
flavour group, and they are normalized so that $\Tr [T^a \, T^b]= \delta_{ab}/2$. Being singlets in color space, for better
readability the summation
over the color index is not shown. The spatially separated two-point correlation functions of these operators
can be defined as 
\begin{equation}\label{eq:2pt}
  C_{{\cal O}}(x_3) =\int dx_0 dx_1 dx_2\, \langle {\cal O}^a (x) {\cal O}^a (0) \rangle\;,
\end{equation}
where no summation over $a$ is understood, and the flavour index has been dropped on the l.h.s.
since $C_{{\cal O}}(x_3)$ does not depend on $a$  when quarks are degenerate. Note that in this case
the disconnected Wick contractions do not contribute. The screening masses are defined as 
\be
m_{{\cal O}} = - \lim_{x_3\rightarrow\infty} \frac{d}{d x_3} \ln\Big[C_{{\cal O}}(x_3)\Big]\; , 
\ee
and they characterize the exponential decrease of the correlation function at large spatial distances.

At low temperatures, due to the chiral anomaly and to the spontaneous breaking of chiral symmetry, the masses resulting from the above
correlation functions are different. When the temperature is large enough, the vector and axial vector screening masses are
expected to become degenerate thanks to the restoration of the non-singlet chiral symmetry. Moreover, at high temperature, the
distribution of the topological charge becomes narrower and narrower~\cite{Giusti:2018cmp}, and only the sector with zero topology
contributes de facto to the
functional integral~\cite{Gross:1980br}, see Refs.~\cite{Giusti:2018cmp,Boccaletti:2020mxu}
and references therein for recent results on this topic. This in practice implies a degeneracy of the non-singlet scalar
and pseudoscalar screening masses as well. 

\subsection{Leading interacting contribution in the effective theory}
The $O(g^2)$ contribution to the non-singlet mesonic screening masses has been computed in the
effective theory~\cite{Hansson:1991kb,Laine:2003bd}. For three massless quarks, the expression reads
\be\label{eq:m_pt}
m^{\rm PT}_{{\cal O}} = 2 \pi T + \frac{g^2_{_{\rm E}}}{3\pi} \big( 1 + 0.93878278\big) =
                    2 \pi T\, (1+0.032739961\cdot g^2)\; , 
\ee
where the first two terms come from the low-energy constant $M$, while the last
one is generated by the interactions~\cite{Laine:2003bd}. Indeed the latter is expected
to receive non-perturbative contributions starting only at $O(g^3)$. In
Eq.~(\ref{eq:m_pt}) the masses are independent of the specific mesonic operator
${\cal O}$ since, as anticipated, spin-dependent effects are expected to appear at $O(g^4)$.

\section{Lattice strategy and setup \label{sec:lattice}}
In order to set up our strategy for studying thermal QCD non-perturbatively up to very high temperatures,
we consider a 4-dimensional lattice with size $L_0$ in the compact (temporal) direction
and extension $L$ along the three spatial directions. As usual, the gauge field is represented
by the link variables $U_\mu(x)\in SU(3)$, while the quark and anti-quark fields are given by
the flavour multiplets $\psi(x)$ and $\psibar (x)$ respectively.

\subsection{Shifted boundary conditions}
The thermal theory is defined by requiring that the fields satisfy shifted boundary
conditions in the compact direction~\cite{Giusti:2011kt,Giusti:2010bb,Giusti:2012yj},
while we set periodic boundary conditions in the spatial directions. The former consist
in shifting the fields by the spatial vector $L_0\, \bsxi$ when crossing the boundary of the
compact direction, with the fermions having in addition the usual sign flip. For the gauge
fields they read
\begin{equation} \label{eq:shift_gluons}
  U_\mu(x_0+L_0,\bsx)= U_\mu(x_0,\bsx-L_0\bsxi)\; ,
  \quad
  U_\mu(x_0,\bsx+\hat{k}L_k)= U_\mu(x_0,\bsx)\; , 
\end{equation}
while those for the quark and the anti-quark fields are given by
\begin{align}
&\psi(x_0+L_0,\bsx)  =  -\psi(x_0,\bsx - L_0\bsxi)\; ,
\quad
& \psi(x_0,\bsx+\hat{k} L_k)  = \psi(x_0,\bsx)\; ,
\nonumber\\
&\psibar(x_0+L_0,\bsx)  =  -\psibar(x_0,\bsx - L_0\bsxi)\;,
\quad
&\psibar(x_0,\bsx+\hat{k} L_k)  = \psibar(x_0,\bsx)\; . \label{eq:shift_quark}
\end{align}
A relativistic thermal field theory in the presence of a shift $\bsxi$ is equivalent to the 
very same theory with usual periodic (anti-periodic for fermions) boundary conditions but with a longer extension of the
compact direction by a factor $\sqrt{1+\vec \xi^2}$~\cite{Giusti:2012yj}, i.e. the standard relation between
the length and the temperature is modified as $T=1/(L_0 \sqrt{1+\vec \xi^2})$. Shifted boundary conditions
represent a very efficient setup to tackle several problems that are otherwise very challenging both from the
theoretical and the numerical viewpoint. A recent example is the EoS of the $SU(3)$ Yang-Mills theory obtained at
the permille level up to very high temperatures~\cite{Giusti:2014ila,Giusti:2016iqr}. The strategy presented in
this paper, when supplemented by shifted boundary conditions, paves the way for the computation of the EoS at large
temperatures in thermal QCD~\cite{DallaBrida:2020gux}. Even if the use of shifted boundary conditions is not
crucial for the calculation of the screening masses, we have chosen to use them with $\vec{\xi}=(1,0,0)$ so as
to share the cost of generating the gauge configurations with that project. The free case computation of the screening
masses reported in appendix~\ref{app:mass_SB}, moreover, indicates that the use of shifted boundary conditions
with $\vec{\xi}=(1,0,0)$ makes discretization effects in the screening masses milder.

\subsection{Renormalization and lines of constant physics}
A hadronic scheme is not a convenient choice to renormalize QCD non-perturbatively when
considering a broad range of temperatures spanning several orders of magnitude. In fact, this
would require to accommodate on a single lattice the temperature and
the hadronic scale which may differ by orders of magnitude, making the numerical
computations prohibitive. A similar problem is encountered when renormalizing QCD
non-perturbatively, and it was solved many years ago by introducing a
step-scaling technique~\cite{Luscher:1991wu,Jansen:1995ck}. 

In order to solve our problem, we build on that knowledge by considering a non-perturbative definition
of the coupling constant, $\bar g^2_{\rm SF}(\mu)$, which can be computed precisely on the lattice for values
of the renormalization scale $\mu$ which span several orders of magnitude. Making a definite
choice, in this section we use the definition based on the Schr\"odinger functional (SF)~\cite{Luscher:1993gh}, but
other choices are available today. In particular later on in the paper we will also consider the
gradient flow (GF) definition~\cite{Fritzsch:2013hda,Brida:2016flw,DallaBrida:2016kgh}. Once $\bar g^2_{\rm SF}(\mu)$ is known in the
continuum limit for $\mu \sim T$ \cite{Brida:2016flw,DallaBrida:2018rfy}, we renormalize thermal QCD
by fixing the value of the renormalized coupling constant at fixed lattice spacing $a$ to be
\be
\bar g^2_{\rm SF}(g_0^2, a\mu) = \bar g^2_{\rm SF}(\mu)\; ,\qquad a\mu\ll 1\;.
\ee
This is the condition that fixes the so-called lines of constant physics, i.e. the dependence of
the bare coupling constant $g_0^2$ on the lattice spacing, for values of $a$ at which the scale $\mu$ and
therefore the temperature $T$ can be easily accommodated. QCD at temperature $T$ can then be simulated
at different values of the lattice spacings, and the continuum limit of the observable of interest
can be taken with confidence. All the technical details on how this procedure is implemented in
practice are given in appendix~\ref{app:IN-OUT}.

\subsection{Lattice setup}
We perform our study at the 12 values of the temperature, $T_0, \ldots$, $T_{11}$,
reported in Table~\ref{tab:M_CL}, covering the range from approximately $1$~GeV up to
about $160$~GeV. For the 9
highest ones, $T_0, \ldots$, $T_8$, gluons are regularized with the Wilson plaquette action in
Eq.~(\ref{eq:SG_W}) of appendix~\ref{app:Lattice}, while for the 3 lowest temperatures, $T_9$, $T_{10}$ and
$T_{11}$, we adopt the tree-level improved gauge action in Eq.~(\ref{eq:SG_I}). The three massless
flavours are always discretized by the $O(a)$-improved Wilson--Dirac operator defined in
appendix~\ref{app:Lattice}.
In order to extrapolate the results to the continuum limit, several lattice spacings are simulated at each
temperature with the extension of the fourth dimension being $L_0/a=4,6,8$ or $10$. The bare coupling and the
critical mass $\mcr$ are fixed at each lattice spacing from the results of
Refs.~\cite{Brida:2016flw,DallaBrida:2016kgh,DallaBrida:2018rfy,Campos:2018ahf} by adopting the strategy outlined above and
explained in details in appendix~\ref{app:IN-OUT}.

\subsection{Finite-volume effects}
As we have discussed in section~\ref{sec:stgy1}, at high temperature the mass gap of the theory
is proportional to $T$ times an appropriate power of the coupling constant. As a consequence,
finite-size effects are proportional to $LT$ times a coefficient that tends to decrease logarithmically
with the temperature, see Refs.~\cite{Meyer:2009kn,Giusti:2012yj} and appendix~\ref{app:finiteV}.
For this reason, the lattices that we consider have rather large spatial directions, i.e. $L/a=288$,
so that $LT$ ranges always from $20$ to $50$. We profit here from the continuous theoretical and
algorithmic progress in the simulation of gauge theories, as well as the steady progress in HPC hardware,
which has made it possible to simulate lattices with a very large number of points. As we will discuss below,
we also always explicitly check that finite-size effects are negligible within the statistical precision
of our observables.

\subsection{Restricting to the zero-topological sector}
At high temperature, the topological charge distribution is expected to be highly peaked
at zero. In particular, in QCD with three light degenerate flavours of mass $m$, the instanton
analysis predicts the topological susceptibility to be proportional to $T^{-b} m^3$ with $b\sim 8$.
The analogous prediction for the Yang--Mills theory has been verified explicitly on the
lattice~\cite{Giusti:2018cmp}. When fermions are introduced, the numerical
computations become significantly more involved and the systematics are still difficult to control.
The simulations done so far, however, are compatible with the $T$-dependence predicted by the
semi-classical analysis~\cite{Bonati:2015vqz,Borsanyi:2016ksw}.
As a result, already at the lower end of the temperature range considered here, $T=1$ GeV, the probability to
encounter a configuration with non-zero topology in our volumes is expected to be several orders of magnitude
smaller than the permille or so. This is even less probable in the limit of massless quarks.
We can therefore restrict our calculations to the sector with  zero topology, and
generate the ensembles of gauge field configurations by a Hybrid Monte Carlo (HMC) as
described in Appendix~\ref{app:OUT}.

\section{Lattice correlation functions and screening masses\label{sec:corfcn}}
After integrating over the fermion fields, the lattice version of the
correlation function in Eq.~(\ref{eq:2pt}) reads 
\begin{equation}\label{eq:2pt_lat}
C_{{\cal O}}(x_3-y_3) =- \frac{a^3}{2} \sum_{x_0,x_1,x_2} \langle
\Tr\Big[\Gamma_{{\cal O}}\, D^{-1}(x,y)\, \Gamma_{{\cal O}}\, \gamma_5 D^{\dagger -1}(x,y) \gamma_5 \Big] 
\rangle\; ,
\end{equation}
where $\Tr$ indicates the trace over the color and spin indices, and the same name as
in the continuum is used since the ambiguity can be resolved
from the context. The quark propagator $D^{-1}(x,y)$ from the source point $y$ to the
sink $x$ is the inverse of the $O(a)$-improved Wilson--Dirac operator defined in Eq.~(\ref{eq:Dirac})
computed at the critical value of the quark mass. At high temperature, the inversion of the lattice Dirac
operator needs to be done with particular care. This is because the lowest Matsubara frequency $\pi T$
provides an infrared cutoff to quark propagation and, as a result, the matrix elements of $D^{-1}(x,y)$
become extremely small when $T\, |x-y| \gg 1$. At those distances a very accurate solution of the Dirac
equation is required, and the brute-force approach of simply implementing higher-precision by requiring a
smaller tolerance is not practicable. We have solved this problem by introducing a distance preconditioning
of the Dirac equation as discussed in Appendix~\ref{app:Dinv}.

The two-point correlation functions for the scalar and pseudoscalar densities and for the vector and axial currents
have been computed on all lattices generated, see Tables~\ref{tab:INparamsWS} and \ref{tab:INparamsGF}. We
report in Tables~\ref{tab:OUTmassesH} and \ref{tab:OUTmassesL} the number of Molecular Dynamics Units (MDUs) after
the thermalization phase of each HMC chain, the number of MDUs skipped between two consecutive independent
configurations, and the number of local sources per configuration on which the Wilson--Dirac operator has been
inverted. The best estimates of $C_{{\cal O}}(x_3)$ on each configuration have been obtained by properly averaging
their values from all local sources and then symmetrizing the correlators with respect to $x_3=L/2$. We carefully
monitored the autocorrelation of the correlators, and we never observed long autocorrelation times with respect
to the number of MDUs skipped between two consecutive measurements.

Within our statistical errors, at all the temperatures that we have investigated,
we observe an excellent agreement between the scalar and pseudoscalar correlators as well
as between the vector and axial ones at intermediate and large distances, see Ref.~\cite{Laudicina:2021gex} for details.
This is a distinctive feature of the restoration of
chiral symmetry which occurs at high temperatures. For this reason, in the rest of the paper we focus our discussion
on the pseudoscalar and the vector correlators only.

\begin{figure}[t!]
  \begin{center}
\includegraphics[width=0.95\textwidth]{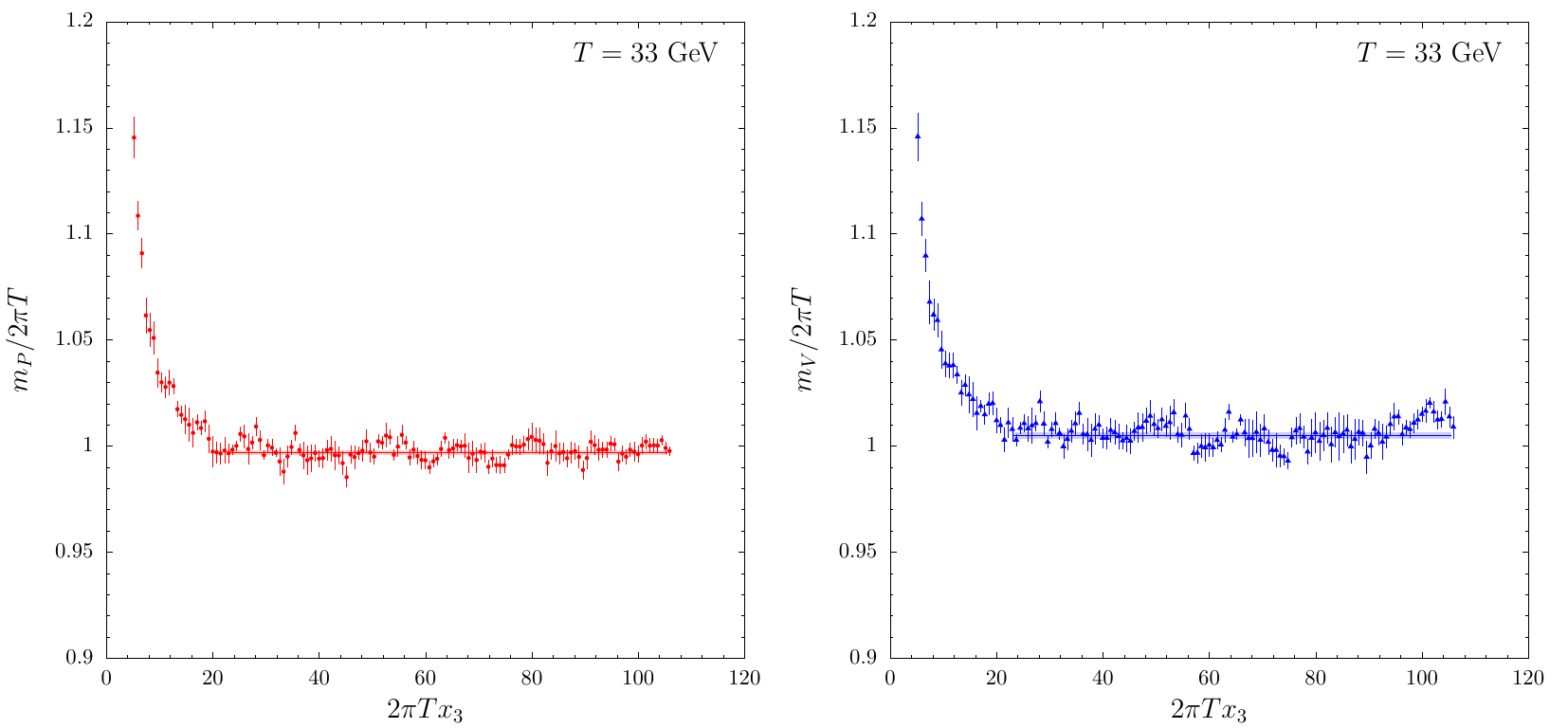}
\caption{Plot of the effective masses, normalized to $2\pi T$, for the pseudoscalar (left) and vector (right) correlators at
the temperature $T_3$ for $L_0/a=6$.\label{fig:plateau}}
\end{center}
\end{figure}

Once the correlation functions in Eq.~(\ref{eq:2pt_lat}) have been computed, effective screening masses
are defined as
\begin{equation}\label{eq:mscrn}
m_{{\cal O}} (x_3)= \frac{1}{a} {\rm arcosh}\Big[\frac{C_{{\cal O}}(x_3+a)+C_{{\cal O}}(x_3-a)}{2\, C_{{\cal O}}(x_3)}\Big]\; . 
\end{equation}
Their values for the pseudoscalar density (left panel) and the vector current (right panel) are shown in
Fig.~\ref{fig:plateau} for $T_3$ and $L_0/a=6$. An analogous behaviour is observed for all
other lattices. We obtain very long plateaux thanks to the fact that we have simulated lattices
with a very large spatial extension, and that there is no signal-to-noise ratio problem
at high temperature. To determine the best estimates of the screening
masses $m_{{\cal O}}$, we start by fitting the symmetrized correlator to a sum of two exponentials from
a minimum value of $x_3/a$ up to the last point available. The minimum value is chosen so as to obtain a
good quality of the two-exponential fit, and at the same time a statistically non-zero contribution from
the sub-leading exponential. From the result of this fit we then estimate 
the minimum value $x_3^{\rm min}/a$ from which the contamination in the effective mass due to the second exponential is
negligible with respect to the statistical precision that we obtain by fitting the effective mass to a constant
from $x_3^{\rm min}/a$ up to the last available point.  We then verify explicitly that a constant value fits well
the effective mass from $x_3^{\rm min}/a$ up to the end of the plateau, and that by increasing $x_3^{\rm min}/a$
by a few units the result of the fit does not change significantly. Examples of results of these fits are represented
in Fig.~\ref{fig:plateau} as straight lines again for $T_3$ and $L_0/a=6$. Our best estimates of the screening masses
are reported in Tables~\ref{tab:OUTmassesH} and \ref{tab:OUTmassesL} for all the lattices simulated. The statistical error is
at most a few permille in all cases. In order to profit from the correlations in our data for reducing the
statistical errors, we also compute $(m_{V}-m_{P})/(2\pi T)$ and report its values in Tables~\ref{tab:OUTmassesH}
and \ref{tab:OUTmassesL} as well. This combination is particularly interesting because it is a measure of the spin-dependent
terms which can be computed very precisely.

We have explicitly checked that finite volume effects are negligible within our statistical errors:
we have generated three more lattices at the highest and at the lowest temperatures for 
the smallest spatial volumes corresponding to $L_0/a=6$, $L_0/a=10$, and $L_0/a=8$ for 
$T_0$, $T_1$, and $T_{11}$ respectively.
These lattices have the same dimensions in the compact and in the $x_3$ directions as those in
Tables~\ref{tab:INparamsWS} and \ref{tab:INparamsGF} but smaller extension in the other
two spatial directions. The screening masses computed on them are in agreement with those
calculated on the larger volume, see appendix~\ref{app:OUT} for the details, and therefore we can
safely assume that our results have negligible finite-volume effects within the statistical precision.

\section{Continuum limit of meson screening masses\label{sect:num_study}}
The results that we have collected at finite lattice spacing have to be extrapolated
to the continuum limit along lines of constant physics. For $O(a)$-improved actions,
the Symanzik effective theory predicts the leading behaviour of the lattice artifacts
to be of order $a^2$. We can accelerate the convergence to the continuum by introducing
the tree-level improved definitions 
\be
m_{{\cal O}} \longrightarrow m_{{\cal O}} -
\Big[m^{\rm free}_{{\cal O}} - 2\pi T\Big]\; , 
\ee
where $m^{\rm free}_{{\cal O}}$ is the mass in the free lattice theory. As 
shown in the appendix~\ref{app:mass_SB}, where the computation is reported, 
the latter is the same for all non-singlet meson masses. From now on we will
consider always the tree-level improved definition of the screening
masses and indicate them with $m_{{\cal O}}$. 
\begin{figure}[t!]
\begin{center}
\includegraphics[width=0.49\textwidth]{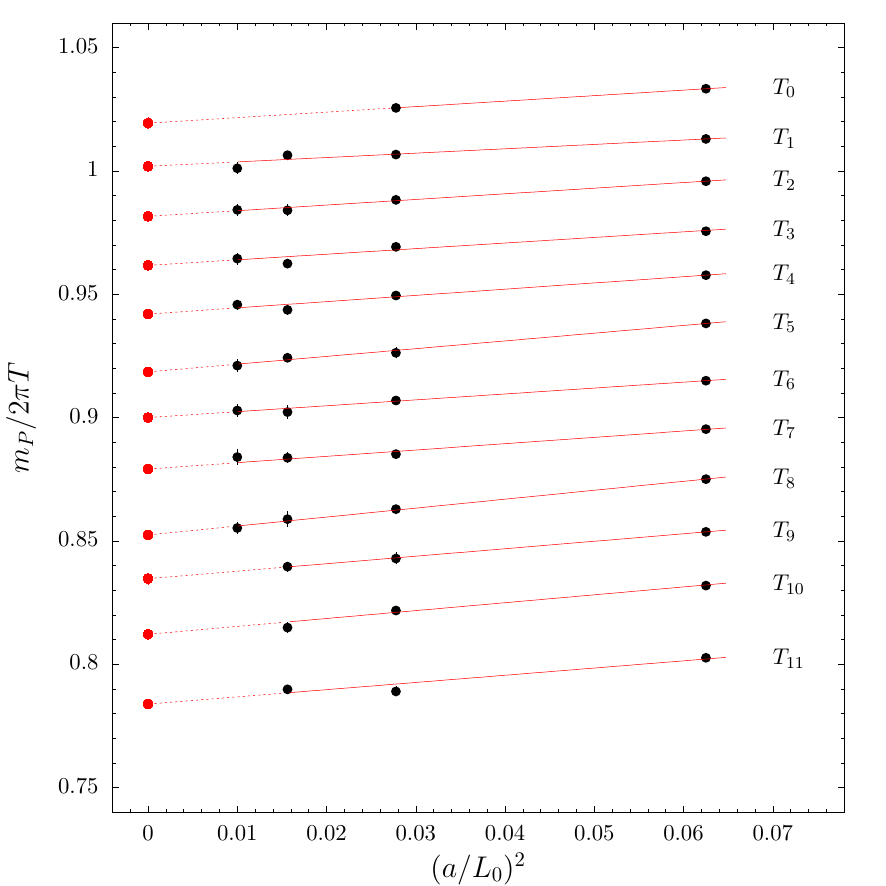}
\includegraphics[width=0.49\textwidth]{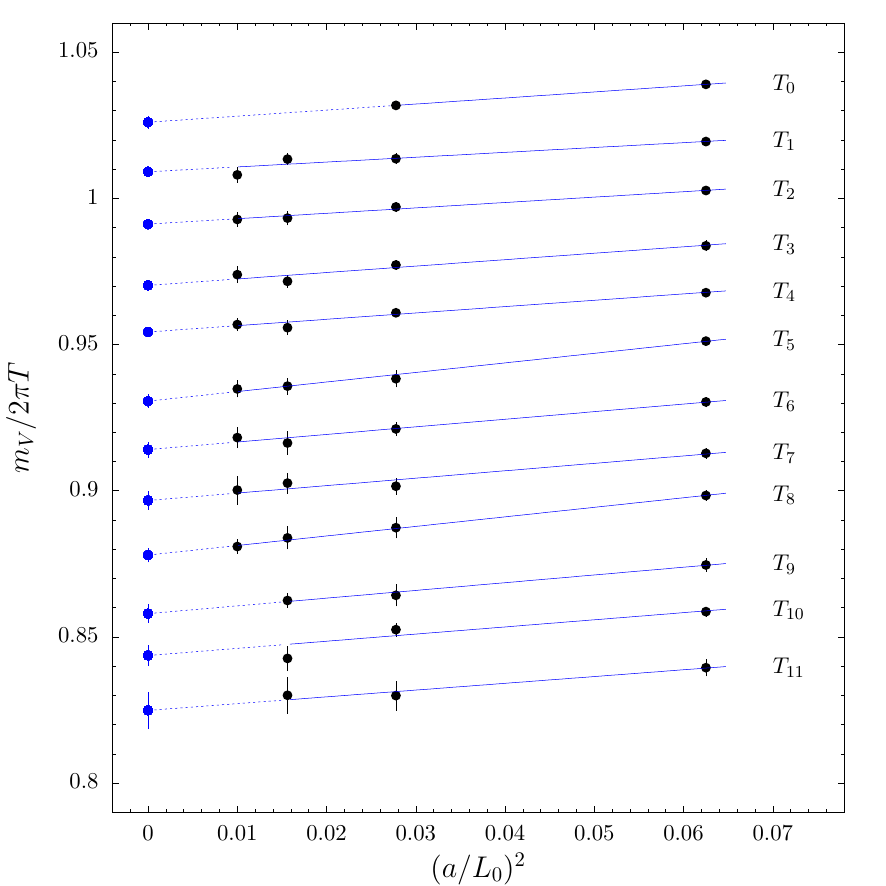}
  \caption{Numerical results for the tree-level improved pseudoscalar (left panel) and vector (right panel) screening masses at finite lattice
  spacing (black dots). The lines in the panels represent the linear extrapolations in $(a/L_0)^2$ to the continuum limit. Each temperature
  is analyzed independently from the others. Data corresponding to $T_i$ ($i=0,\dots,11$) are shifted downward by $0.02\times i$ for better readability.
\label{fig:PV_extrCL}}
\end{center}
\end{figure}

\begin{wrapfigure}{r}{7.5cm}
  \centering
\includegraphics[width=0.49\textwidth]{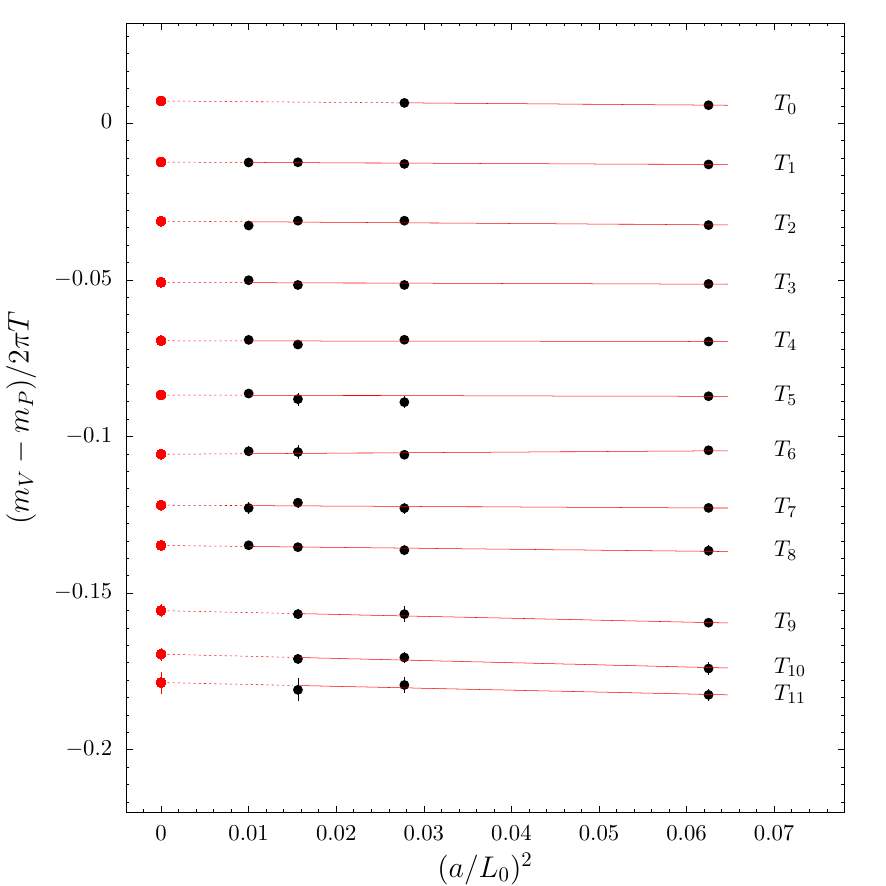}      
  \caption{As in Fig.~\ref{fig:PV_extrCL} but for the mass difference.\label{fig:DPV_extrCL}}
\vspace{-0.5cm}

\end{wrapfigure}

All data for the improved pseudoscalar (left panel) and vector (right panel) screening masses are
represented in Fig.~\ref{fig:PV_extrCL} where, in order to improve the readability, data corresponding
to $T_i$ ($i=0,\dots,11$) are shifted downward by $0.02\times i$. The analogous plot for $(m_{V}-m_{P})$
is shown in Fig.~\ref{fig:DPV_extrCL}. At each temperature, lattice artifacts  are well described by a single
correction proportional to $(a/L_0)^2$. Indeed by fitting each data set linearly in $(a/L_0)^2$, the values of
$\chi^2/{\rm dof}$ are all around $1$ with just a few outliers which, however, are not surprising given the 
large
amount of data and fits. The results of the fits are shown in the plots of Figs.~\ref{fig:PV_extrCL} and \ref{fig:DPV_extrCL} as straight lines. For the
mass difference, the coefficient of $(a/L_0)^2$ is found to be compatible with zero at all temperatures.
We take the continuum limit values from these fits as our best results for the non-singlet meson screening masses and their difference. They
are reported in Table~\ref{tab:M_CL} for all the 12 temperatures considered. 
As a further check of the extrapolations, we have fitted the
data by excluding the coarsest lattice spacing, i.e. $L_0/a=4$, for the temperatures $T_1,\ldots, T_{8}$ for which
we have $4$ data points. The intercepts are in excellent agreement with those of the
previous fits, albeit with a slightly larger error. For the same sets of data, we have also attempted to include in the fit a
$(a/L_0)^2\ln(a/L_0)$ or a $(a/L_0)^3$ term. The resulting coefficients are compatible with zero. Given the high quality
of the fits and of the data, it is not necessary to model the temperature dependence of the discretization effects so as
to perform a global fit of the data.
\begin{table}[th!]
\centering
\begin{tabular}{|c|c|c|c|c|}
\hline
 & & & & \\[-0.25cm]
$T$ & $T({\rm GeV})$  & $\displaystyle \frac{m_{P}}{2\pi T} $ & $\displaystyle\frac{m_{V}}{2\pi T}$ &
 $\displaystyle\frac{(m_{V}-m_{P})}{2\pi T}$\\[-0.25cm]
  & & & & \\
\hline
$T_0$ & 164.6(5.6) & 1.0194(25)  & 1.0261(23)   & 0.0071(7) \\
$T_1$ &  82.3(2.8) & 1.0219(15)  & 1.0291(18)   & 0.0076(4) \\
$T_2$ &  51.4(1.7) & 1.0216(16)  & 1.0312(18)   & 0.0087(4)\\
$T_3$ &  32.8(1.0) & 1.0217(15)  & 1.0302(19)   & 0.0092(6)\\
$T_4$ &  20.63(63) & 1.0220(15)  & 1.0343(17)   & 0.0105(6)\\
$T_5$ &  12.77(37) & 1.0185(18)  & 1.0306(24)   & 0.0132(10)\\
$T_6$ &  8.03(22)  & 1.0200(18)  & 1.0341(28)   & 0.0143(13)\\
$T_7$ &  4.91(13)  & 1.0192(18)  & 1.037(3)   & 0.0181(14)\\
$T_8$ &  3.040(78) & 1.0124(18)  & 1.0380(25)   & 0.0252(13)\\
$T_9$ &  2.833(68) & 1.0147(24)  & 1.038(3)   & 0.0244(20)\\
$T_{10}$& 1.821(39) & 1.0122(18)  & 1.044(4)   & 0.0305(20)\\
$T_{11}$& 1.167(23) & 1.0039(20)  & 1.045(6)   & 0.041(4)\\
\hline
\end{tabular}
\caption{Best results for the pseudoscalar, $m_{P}$, and the vector, $m_{V}$, non-singlet screening masses in the
continuum limit together with their difference.\label{tab:M_CL}}
\end{table}

\vspace{-0.25cm}

\section{Discussion and interpretation of the results\label{sec:disc}}
The main results of this paper are the non-singlet meson screening masses reported in
Table~\ref{tab:M_CL}. They have been computed in a wide temperature range starting from
$T\!\sim\! 1$~GeV up to $160$~GeV or so with a precision of a few permille.

The first observation is that, as anticipated in section~\ref{sec:corfcn},
within our rather small statistical errors we
find an excellent agreement between the scalar and pseudoscalar masses and the vector
and axial ones. This is a clear manifestation of the restoration of chiral symmetry occurring
at high temperature. For this reason we do not show explicitly the results for the
other two channels, and we focus on the pseudoscalar and vector masses.

A second observation is that the bulk of the non-singlet meson screening masses is given by the free-theory value,
$2\pi T$, plus a few percent positive contribution over the entire range of temperatures explored.

Thanks to the precision of our results, we can scrutinize in detail the temperature
dependence induced by the non-trivial dynamics. We introduce the function $\hat g^2 (T)$ defined as 
\begin{equation}\label{eq:gmu}
  \frac{1}{\hat g^2(T)} \equiv \frac{9}{8\pi^2} \ln  \frac{2\pi T}{\Lambda_{\MSbar}} 
  + \frac{4}{9 \pi^2} \ln \left( 2 \ln  \frac{2 \pi T}{\Lambda_{\MSbar}}  \right)\; , 
\end{equation}
where $\Lambda_{\MSbar} = 341$~MeV is taken from Ref.~\cite{Bruno:2017gxd}. It corresponds
to the 2-loop definition of the strong coupling constant in the
$\MSbar$ scheme at the renormalization scale $\mu=2\pi T$. For our purposes, however, this is just a function
of the temperature $T$, suggested by the effective theory analysis, that we use to analyze our results\footnote{One could also use
a non-perturbative definition of the coupling constant, such as $\bar g^2_{\rm SF}$. In this
case, however, comparing our data with the analytic results in the literature would be more involved.}. The crucial
point is the leading logarithmic dependence on $T$.
\begin{figure}[t!]
\begin{center}
\includegraphics[width=0.49\textwidth]{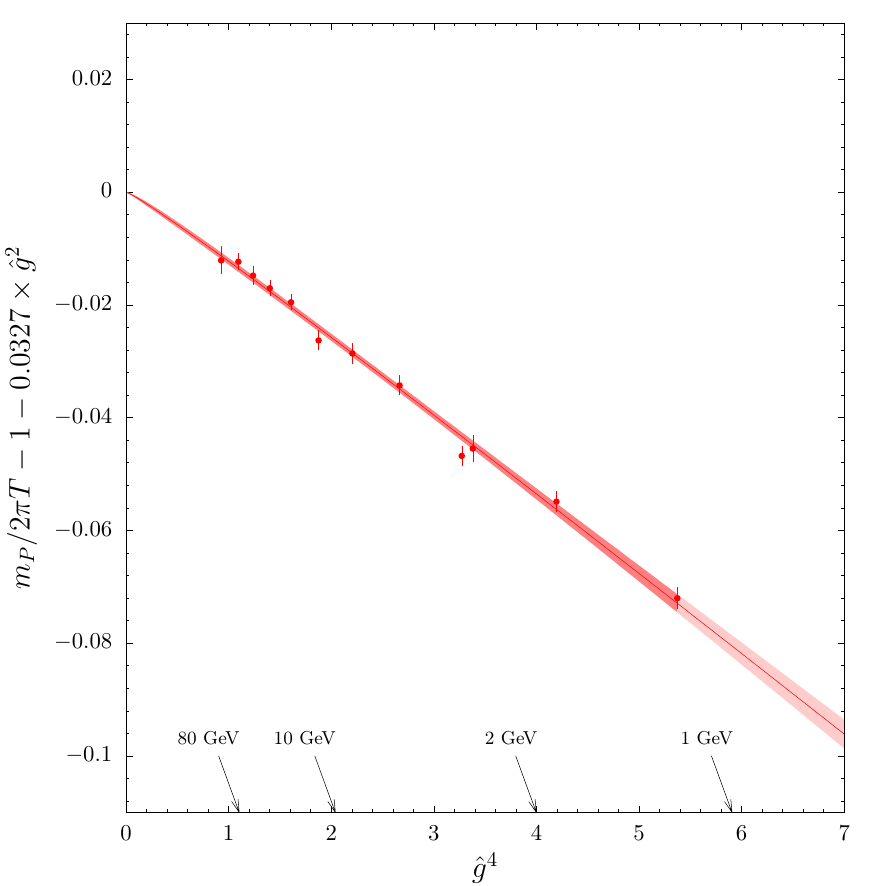}
\includegraphics[width=0.49\textwidth]{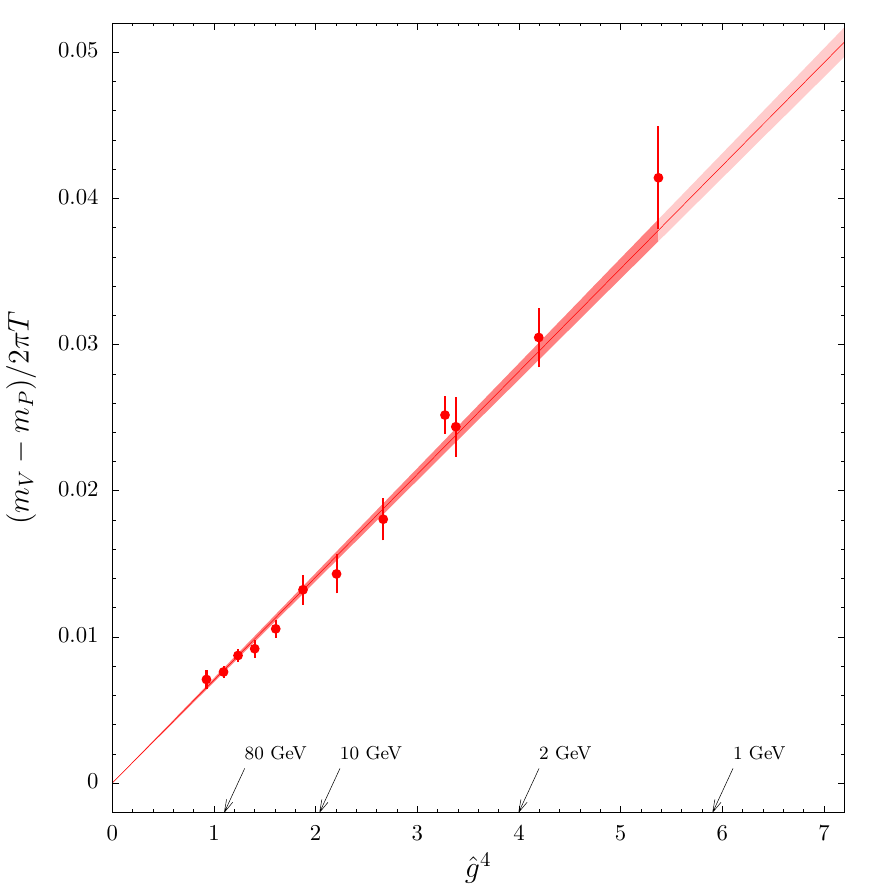}
  \caption{Left: the pseudoscalar mass, normalized to $2\pi T$, subtracted of the analytically known contributions versus $\hat g^4$.
           Right: the vector-pseudoscalar mass difference, normalized to $2\pi T$, versus $\hat g^4$. Red bands
           represent the best fits of the data as explained in the text.
\label{fig:PV_extr_g^p}}
\end{center}
\end{figure}

\subsection{Pseudoscalar mass}
We start our analysis by fitting the pseudoscalar mass in the third column of Table~\ref{tab:M_CL}
to a quartic polynomial in $\hat g$. The intercept turns out to be compatible with $1$, as predicted
by the free theory, within a large error. We have thus enforced it to the free-theory value, $p_0=1$, and 
we have fitted again the data. The coefficient of the $\hat g^2$
term turns out to be compatible with the theoretical expectation in Eq.~(\ref{eq:m_pt})
within again a large uncertainty. We have thus fixed also this coefficient to its analytical
value, $p_2=0.032739961$, and we have performed again the quartic fit of the form
\be\label{eq:quartic}
\frac{m_{P}}{2\pi T} = p_0 + p_2\, \hat g^2 + p_3\, \hat g^3 + p_4\, \hat g^4\; .  
\ee
As a result, for the fit parameters we obtain $p_3=0.0038(22)$, $p_4=-0.0161(17)$ and 
${\rm cov}(p_3,p_4)/[\sigma(p_3) \sigma(p_4)]=-1.0$
with the excellent value of $\chi^2/{\rm dof}=0.75$. The quality of the fit can be appreciated in the left plot of
Fig.~\ref{fig:PV_extr_g^p}, where $m_{P}/(2\pi T)$ - subtracted of the analytically
known contributions - is shown as a function of $\hat g^4$ together with the best fit to Eq.~(\ref{eq:quartic}).
If the cubic coefficient is enforced to vanish, i.e. $p_3=0$, the fit returns $p_4=-0.01323(20)$ with
again an excellent value of $\chi^2/{\rm dof}=0.96$. The subtracted data lie on a straight
line over two orders of magnitude in the temperature. The polynomial in Eq.~(\ref{eq:quartic}) is 
our best parameterization of the results over the entire range of temperatures explored. 

The quartic term is necessary to explain the data over the entire
temperature range. In particular at the electroweak scale or so, it is still
approximately half of the total contribution due to the interactions. 
Notice that the sign of the quartic term is negative,
opposite to the one of the quadratic contribution, and the magnitude turns out to be approximately
$2$--$3$ times smaller than $p_2$. When the data are plotted as a function of $\hat g^2$, the quartic contribution
competes with the quadratic one to bend down the pseudoscalar mass as shown in Fig.~\ref{fig:massCL}.
Toward the lower end of the range, the competition between this term and
the leading one results in an effective slope of opposite sign with respect to the analytically
known one. At $T \sim 1$~GeV, the various terms cancel each other and the mass turns out
to be very close to free-value $2\pi T$. 

\subsection{Vector mass}
The mass difference $(m_{V}\!-\!m_{P})/(2\pi T)$ is an interesting quantity to
investigate the magnitude of the spin-dependent contributions. We plot our results for
this quantity (last column of Table~\ref{tab:M_CL}) as a function of $\hat g^4$ on the
right panel of Fig.~\ref{fig:PV_extr_g^p}. 
The data turn out to lie on a straight line with a vanishing intercept. By fitting them to
\be\label{eq:spindep}
\frac{(m_{V} - m_{P})}{2\pi T} = s_4\, \hat g^4\; ,  
\ee
we obtain $s_4=0.00704(14)$ with $\chi^2/{\rm dof}=0.79$. It turns out that the spin-dependent
contribution can be parameterized by a single $O(\hat g^4)$ term in the entire range of
temperatures explored. Furthermore, it remains clearly visible up to the highest temperature, where
the pseudoscalar and the vector masses are still significantly different within our
numerical precision,
\begin{wrapfigure}{r}{7.5cm}
  \centering
\includegraphics[width=0.49\textwidth]{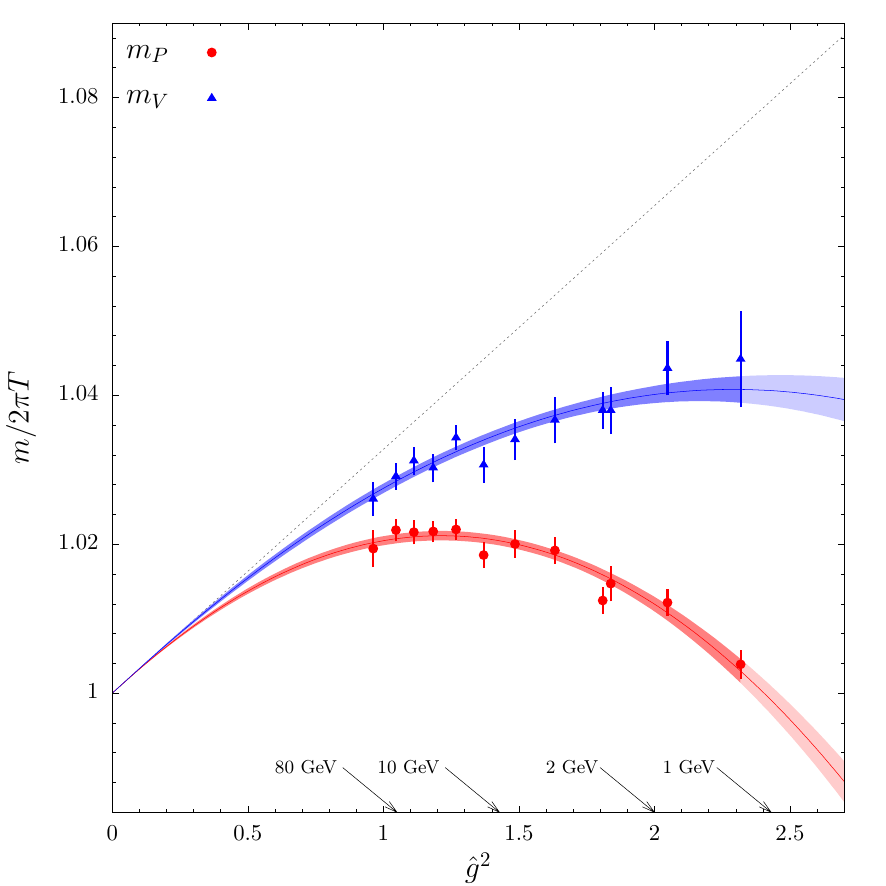}      
\caption{Pseudoscalar (red) and vector (blue) screening masses versus $\hat g^2$. The bands
         represent the best fits in Eqs.~(\ref{eq:quartic}) and (\ref{eq:mvfit}), while the
         dashed line is the analytically known contribution.\label{fig:massCL}}
\end{wrapfigure}
see Fig.~\ref{fig:massCL}. The best polynomial that parameterizes our results for the vector mass (fourth column
of Table~\ref{tab:M_CL}) is therefore 
\be\label{eq:mvfit}
\frac{m_{V}}{2\pi T} = p_0 + p_2\, \hat g^2 + p_3\, \hat g^3 + (p_4 + s_4) \, \hat g^4\; ,
\ee
where $p_0,\ldots, p_4$ are those in Eq.~(\ref{eq:quartic}) while $s_4$ is
taken from Eq.~(\ref{eq:spindep}). The covariances of the coefficients $p_3$ and
$p_4$ with $s_4$ are ${\rm cov}(p_3,s_4)/[\sigma(p_3) \sigma(s_4)]=0.08$ and
${\rm cov}(p_4,s_4)/[\sigma(p_4) \sigma(s_4)]=-0.07$. 

As shown in Fig.~\ref{fig:massCL}, the quartic contribution is necessary to explain the
data over the entire temperature range. In particular at the electroweak
scale, it is still approximately $15\%$ of the total contribution due to the interactions. 
Also for the vector mass, the coefficient of the quartic term in Eq.~(\ref{eq:mvfit})
has an opposite sign with respect to $p_2$, but it is approximately half of the analogous
one for the pseudoscalar. When the mass is plotted as a function of $\hat g^2$, see
Fig.~\ref{fig:massCL}, the quartic contribution competes with the quadratic one but is not
large enough to push down the vector mass, at least in the range considered.  At the lower end
of our range, $T \sim 1$~GeV, it is the spin-dependent term that is responsible for the
deviation of the vector mass from $2 \pi T$, given the
cancellation among the other terms.\\[0.125cm]

In the literature, non-perturbative computations of these masses are available only up to
temperatures of $1$~GeV or so~\cite{Bazavov:2019www}. Even if these results have been obtained
at the physical values of the quark masses or close by, they are in agreement with ours at those
temperatures within the rather large errors. In fact, the relevance of the quark masses at that
temperature is very mild, and it becomes quickly negligible as the temperature increases.
The pattern of different contributions
that we have just discussed, however, explains why it has been difficult in the past to match
non-perturbative lattice results at $T \lesssim 1$~GeV with the expected analytic behaviour at
asymptotically high temperatures. Indeed the apparently small 2-4\% effect in the screening
masses induced by the interactions among quarks and gluons encodes a lot of interesting non-trivial
information about the dynamics of the plasma. When the corresponding non-perturbative
computations in the three-dimensional effective theory will become available, the matching with
the results presented here will allow to shed light on the origin of the various terms, and to
verify non-perturbatively the effective theory paradigm over several orders of magnitude in the
temperature.

\section{Conclusions and outlook \label{sec:conc}}
The continuous theoretical and algorithmic progress in the simulation of gauge theories, as well as the steady
progress in HPC hardware, has made it possible to simulate lattices with a very large number of
points. This has opened to the possibility of studying thermal gauge theories non-perturbatively at very high
temperature. Here we have profited from this progress to simulate, for the first time, QCD with three massless
flavours at temperatures ranging from 1 GeV up to the electroweak scale and above. We have
renormalized the theory by imposing the value of the strong coupling constant defined non-perturbatively
in a finite-volume renormalization scheme. The
entire strategy has been implemented by discretizing fermions with the $O(a)$-improved Wilson--Dirac operator. This is a
theoretically-sound regularization, not only simple to simulate, but which enjoys de-facto automatic
$O(a)$-improvement at high temperature~\cite{DallaBrida:2020gux}.

In the very high temperature regime, the non-singlet meson screening masses
are perhaps the simplest computations to start with.
For QCD with three massless flavours, we observe an excellent agreement within our
statistical errors between the scalar and pseudoscalar masses as well as between the vector and axial ones.
This is a clear manifestation of the restoration of chiral symmetry for temperatures from
$T\!\sim\! 1$~GeV up to $160$~GeV. Our best results for the pseudoscalar and
vector masses are reported in Table~\ref{tab:M_CL}, they are parameterized in Eqs.~(\ref{eq:quartic}) and (\ref{eq:mvfit}),
and they are shown in Fig.~\ref{fig:massCL}. In the entire range explored,
the meson screening masses deviate from the free theory result, $2 \pi T$, by at most a
few percent. These deviations, however, cannot be explained by the known leading term
in the coupling constant up to the highest temperature, where other contributions are
still very significant. The latter bend down the pseudoscalar mass to the point that
the effective slope in $\hat g^2$, at the lower temperatures, is of opposite sign with
respect to the analytically known one. The spin-dependent contributions are very well parameterized, within
our statistical errors, by a single $O(\hat g^4)$ term in the entire range of temperature explored. At low temperatures,
and in particular at $T \sim 1$~GeV, this term is responsible for the deviation of the vector mass
from $2 \pi T$. It remains clearly visible up to the highest temperature, where the pseudoscalar and the
vector masses are still significantly different within our numerical precision.

The pattern of different contributions that we have found explains why it has been
difficult in the past to match non-perturbative lattice results
at $T \lesssim 1$~GeV with the known  analytic behaviour at asymptotically high temperatures.
From a more theoretical point of view, when the corresponding non-perturbative computations in the three-dimensional
effective theory will become available, the matching with the results presented here will allow to shed light on
the origin of the various terms, and to verify non-perturbatively the effective theory paradigm. At the same time,
the possibility of studying QCD at high temperatures by Monte Carlo simulations, makes the perturbative results less
compelling especially if higher and higher orders must be taken into account.  

The strategy proposed here clears the way to compute many other interesting properties of thermal QCD in the
high temperature regime. Indeed this work is part of a larger effort which aims at computing the EoS non-perturbatively
up to the electroweak scale or so.

\section*{Acknowledgement}
We wish to thank Mikko Laine for correspondence on the expansion in the strong coupling of the screening masses.
We acknowledge PRACE for awarding us access to
the HPC system MareNostrum4 at the Barcelona Supercomputing Center
(Proposals n. 2018194651 and 2021240051) where most of the numerical results presented in this paper have been produced. We also thanks
CINECA for providing us with computer-time on Marconi (CINECA- INFN, CINECA-Bicocca agreements, ISCRA B projects
HP10BF2OQT and HP10B1TWRR).
The R\&D has been carried out on the PC clusters Wilson and Knuth at Milano-Bicocca. We thank all these institutions for
the technical support. Finally we acknowledge partial support by the INFN project ``High performance data network''.
T. H. has been supported by UK STFC CG ST/P000630/1.

\appendix

\section{Lattice actions \label{app:Lattice}}
The QCD lattice action is $S=S_G+S_F$ where $S_G$ and $S_F$ are the pure gauge and the fermionic
parts respectively. In this paper we use both the Wilson plaquette action, $S_G^{(W)}$, and the tree-level Symanzik
improved action, $S_G^{(I)}$, for the pure gauge sector. The former is defined as~\cite{Wilson:1974sk}
\begin{equation}\label{eq:SG_W}
S_G^{(W)}= {1\over g_0^2} \sum_x \sum_{\mu,\nu} {\rm Re}\,\Tr\big[1\!\! 1-U_{\mu\nu}(x)\big]  
\end{equation} 
where the plaquette field is
\begin{equation}
  U_{\mu\nu}(x)=U_\mu(x)\, U_\nu(x+ a \hat{\mu})\, U_\mu^\dag(x+ a \hat{\nu})\, U_\nu^\dag(x)\; ,
\end{equation}
and $\hat{\mu},\hat{\nu}$ are unit vectors oriented along the directions $\mu,\nu$ respectively.
The tree-level Symanzik improved action is defined as~\cite{Luscher:1984xn}
\begin{equation}\label{eq:SG_I}
  S_G^{(I)}= {1\over g_0^2} \sum_x \sum_{\mu,\nu} {\rm Re}\,
\Big\{
\frac{5}{3}  \Tr\big[1\!\! 1-U_{\mu\nu}(x)\big]  
-\frac{1}{12} \Tr\big[1\!\! 1-\widetilde U_{\mu\nu}(x)\big]  
\Big\}
\end{equation} 
where $\widetilde U_{\mu\nu}$ is a rectangular two-plaquette field defined as 
\begin{equation}
  \widetilde U_{\mu\nu}(x)=
  U_\mu(x)\, U_\mu(x+ a \hat{\mu}) \, U_\nu(x+ 2 a \hat{\mu}) \,  U_\mu^\dag(x+ a \hat{\mu} + a \hat{\nu}) \, U_\mu^\dag(x+ a \hat{\nu}) \, U_\nu^\dag(x)
    \; .
\end{equation}
The fermionic part of the action is 
\begin{equation}\label{eq:SF}
S_F=a^4\sum_x \psibar(x) (D+M_0)\psi(x)
\end{equation}
where $M_0$ is the bare quark mass matrix and $D$ is the lattice Dirac operator for which we consider the
$O(a)$-improved definition~\cite{Wilson:1975hf,Sheikholeslami:1985ij}
\begin{equation}
  \label{eq:Dirac}
  D=D_{\rm w} + a D_{\rm sw}\; . 
\end{equation}
The first term $D_{\rm w}$ is the massless Wilson-Dirac operator given by
\begin{equation}
  D_{\rm w} =
  \frac{1}{2}\big\{\dirac\mu(\nabstar\mu+\nab\mu)-a\nabstar\mu \nab\mu\big\}\; ,
\end{equation}
where $\nabla_\mu^*,\nabla_\mu$ are covariant lattice derivatives that act on the quark fields as follows
\begin{align}
a \nab\mu \psi(x) & =  U_\mu(x)\psi(x+ a \hat{\mu})-\psi(x)\; ,\nonumber \\[0.25cm]
a \nabstar\mu \psi(x) & = \psi(x) - U^\dag_\mu(x- a \hat{\mu})\psi(x - a \hat{\mu})\; .
\end{align} 
The second term $D_{\rm sw}$ is the Sheikholeslami-Wohlert operator 
\begin{equation}\label{eq:DiracSW}
  D_{\rm sw}\psi(x) = c_{\rm sw}(g_0) \frac{1}{4}
  \sigma_{\mu\nu} \widehat F_{\mu\nu}(x)\psi(x)\; 
\end{equation}
where $\sigma_{\mu\nu}=\frac{i}{2}[\dirac\mu,\dirac\nu]$. The field $\widehat F_{\mu\nu}(x)$ is
the clover discretization of the field strength tensor which is given by
\begin{equation}
\label{eq:CloverFmunu}
\widehat  F_{\mu\nu}(x) = \frac{i}{8a^2}\big\{Q_{\mu\nu}(x)-Q_{\nu\mu}(x)\big\}\; ,
\end{equation}
with
\begin{equation}
\begin{split}
	Q_{\mu\nu}(x) &= U_\mu(x)U_\nu(x+a \hat{\mu})U^\dag_\mu(x+a \hat{\nu})U^\dag_\nu(x)\\
	&+ U_\nu(x)U_\mu^\dag(x-a \hat{\mu}+a \hat{\nu})U^\dag_\nu(x-a \hat{\mu})U_\mu(x-a \hat{\mu})\\
	&+ U_\mu^\dag(x-a \hat{\mu})U_\nu^\dag(x-a \hat{\mu}-a \hat{\nu})U_\mu(x-a \hat{\mu}-a \hat{\nu})U_\nu(x-a \hat{\nu})\\
	&+ U_\nu^\dag(x-a \hat{\nu})U_\mu(x-a \hat{\nu})U_\nu(x+a \hat{\mu}-a \hat{\nu})U_\mu^\dag(x)\; .
\end{split}
\end{equation}
By a proper non-perturbative tuning of the coefficient $c_{\rm sw}(g_0)$, all $O(a)$ discretization effects
generated by the action in on-shell correlation functions can be removed~\cite{Sheikholeslami:1985ij,Luscher:1996sc}.
For the Wilson plaquette action this is achieved by fixing $c_{\rm sw}(g_0)$ to \cite{Yamada:2004ja} 
\begin{equation}\label{eq:cSW-W}
  c_{\rm sw}^{(W)} (g_0)=\frac
  {1 - 0.194785\, g_0^2 - 0.110781\, g_0^4 - 0.0230239\, g_0^6 + 0.137401\, g_0^8}
  {1-0.460685\, g_0^2}\; ,
\end{equation}
while for the tree-level Symanzik improved gauge action the analogous expression is \cite{Bulava:2013cta} 
\begin{equation}\label{eq:cSW-I}
  c_{\rm sw}^{(I)} (g_0)=\frac
  {1 - 0.1921\, g_0^2 - 0.1378\, g_0^4 + 0.0717\, g_0^6}
  {1 - 0.3881\, g_0^2}\; . 
\end{equation}

\section{Temperature values and lines of constant physics\label{app:IN-OUT}}
In this appendix we discuss in detail how the 12 temperatures $T_0, \ldots$, $T_{11}$ have been chosen,
and how for each temperature the various lattice spacings and the corresponding bare parameters have been
fixed so as to define lines of constant physics.

Either for quarks or gluons we remind that shifted boundary conditions, always with $\bsxi=(1,0,0)$, have been
enforced in the compact direction so that $T={1/(\sqrt{2} L_0)}$. The temperature values $T_0, \ldots$, $T_8$ and
$T_9, \ldots$, $T_{11}$  have then been fixed by specifying the values of the Schr\"odinger functional (SF) and
the gradient flow (GF) finite-volume couplings respectively by using the results of
Refs.~\cite{Brida:2016flw,DallaBrida:2016kgh,DallaBrida:2018rfy,Campos:2018ahf}. 
\begin{table}[t!]
	\centering
	\small
	\begin{tabular}{|c|c|c|}
		\hline
		$T$ &  $\bar{g}^2_{\rm SF}(\mu=T\sqrt{2})$ & $T$ (GeV) \\
		\hline
		$T_0$ &   $-$    &  164.6(5.6) \\
		$T_1$ &  1.11000 &  82.3(2.8)  \\
		$T_2$ &  1.18446 &  51.4(1.7)  \\
		$T_3$ &  1.26569 &  32.8(1.0)  \\
		$T_4$ &  1.3627  &  20.63(63)  \\   
		$T_5$ &  1.4808  &  12.77(37)  \\
		$T_6$ &  1.6173  &  8.03(22)   \\
		$T_7$ &  1.7943  &  4.91(13)   \\
		$T_8$ &  2.0120  &  3.040(78)  \\
		\hline
	\end{tabular}
	\caption{Values of the SF couplings corresponding to the 
		lines of constant physical temperature that we consider.}
	\label{tab:RenCouplingValuesSF}	
\end{table}

\subsection{High temperatures}
The temperature values $T_0,\ldots,T_8$ are fixed from the results in
Refs.~\cite{Brida:2016flw,DallaBrida:2018rfy,Campos:2018ahf}
by imposing the relation
\begin{equation}
	T={1\over L_0\sqrt{2}}={\mu\over \sqrt{2}}\,,
\end{equation}
where $\mu$ is the renormalization scale of the Schr\"odinger functional (SF) coupling
$\bar{g}^2_{\rm SF}(\mu)$ determined in a box with linear extension $L^{\rm SF}_0 = 1/\mu$
and SF boundary conditions enforced, i.e. $L_0=L^{\rm SF}_0$. From Ref.~\cite{Bruno:2017gxd}
we obtain  
\begin{equation}
	\label{eq:ValueMu0}
	\bar{g}^2_{\rm SF}(\mu_0)=2.0120 
	\quad
	\Rightarrow
	\quad
	\mu_0=4.30(11)\,{\rm GeV}=T_8\sqrt{2}\,,
\end{equation}
where the contribution from the charm and bottom quarks can be safely neglected given the current level
of precision on the combination of the pion and kaon decay constants used to fix the overall
scale~\cite{Athenodorou:2018wpk}, see Ref.~\cite{Bruno:2016plf} for more details. Given $T_8$, the higher
values of the temperature can be inferred through the relation
\begin{equation}
	\label{eq:MuRatioSF}
	\ln\bigg({\mu\over\mu_0}\bigg)=\int_{\bar{g}_{\rm SF}(\mu_0)}^{\bar{g}_{\rm SF}(\mu)}{{\rm d }g\over \beta_{\rm SF}(g)}\,,
\end{equation}
which readily follows from integrating the definition of the $\beta$-function. By using the results of
Ref.~\cite{DallaBrida:2018rfy}, the non-perturbative $\beta$-function of the SF coupling can be parameterized over
the range of couplings of interest as (cf.~Eq.~(2.34) of Ref.~\cite{Campos:2018ahf})
\begin{equation}
	\beta_{\rm SF}(\bar{g})=-\bar{g}^3\sum_{n=0}^3b_n\bar{g}^{2n},
	\qquad
	\bar{g}^2\in[0,2.45]\,,
\end{equation}
with $b_0,b_1,b_2$ being the perturbative coefficients of the SF $\beta$-function (for $N_{\rm f}=3$) 
\begin{gather*}
	(4\pi)b_0={9\over 4\pi}\,, 
	\qquad 
	(4\pi)^2b_1={4\over \pi^2}\,,
	\qquad
	(4\pi)^3b_2=-0.064(27)\,,
\end{gather*}
while $b_3$ is an effective higher-order contribution extracted from the 
non-perturbative data
\begin{equation}
	(4\pi)^4b_3^{\rm eff}=4(3)\,.
\end{equation}	
Given this representation, we integrated Eq.~(\ref{eq:MuRatioSF}) numerically
using the result for $\mu_0$ in Eq.~(\ref{eq:ValueMu0}), and we obtained the values for
the temperatures reported in Table \ref{tab:RenCouplingValuesSF}.
\begin{table}[hbtp!]
	\small
	\centering
	\begin{tabular}{|l|l|l|}
		\hline
		$L_0/a$ & $\delta a\mcr[(0)]$ & $\delta a\mcr[(1)]$ \\
                \hline
		$  4$ & $-0.0015131$         & $0.0120930$          \\
		$  6$ & $-0.0006384$         & $0.0008250$          \\
		$  8$ & $-0.0003209$         & $0.0001878$          \\
		$ 10$ & $-0.0001835$         & $0.0000751$          \\
		$ 12$ & $-0.0001145$         & $0.0000403$          \\
		$ 16$ & $-0.0000531$         & $0.0000168$          \\
		\hline
	\end{tabular}
	\caption{\label{tab:damcrSF}\small%
		Tree-level and one-loop cutoff effects for the critical mass in
		the SF for setup~A with background gauge field, $\theta=\pi/5$ and
		improvement coefficients as specified in Ref.~\cite{Kurth:2002rz}.
                Note that the one-loop
		coefficient depends on the number of flavours, $\delta a\mcr[(1)]
		= \delta a\mcr[(1,0)] + \delta a\mcr[(1,1)]N_{\rm f}$, with numerical
		values taken from~\cite{Kurth:2002rz}.}
\end{table}

\subsubsection{Bare parameters}
For the 9 highest temperatures $T_0, \ldots$, $T_8$, we opted for the Wilson plaquette
action in Eq.~(\ref{eq:SG_W}). This allows us to fix the bare parameters along the lines
of constant physics by exploiting the known results for the SF coupling computed de
facto at the critical mass. For $L_0/a=6,8,10$ they are given in Table~3 of
Ref.~\cite{DallaBrida:2018rfy}, while those for $L_0/a=4$ were given to us by the
authors of that reference as a private communication. For each value of $L_0/a$,
we can determine the values of $\beta$ at which the SF coupling has the prescribed
values reported in Table \ref{tab:RenCouplingValuesSF} by fitting
$\bar{g}^2_{\rm SF}(\mu)$ to the functional form\footnote{The results for $\bar{g}^2_{\rm SF}$ from
Ref.~\cite{DallaBrida:2018rfy} come with an error which includes both statistical and systematic
uncertainties. The latter is an estimate for the remaining $O(a g_0^8)$ effects stemming from the SF
boundary counter-terms after the known perturbative improvement is implemented. We have explicitly checked
that, once propagated to the screening masses, these errors are negligible within the statistical uncertainties.
We can therefore safely assume that the screening masses are free from $O(a)$ contaminations deriving from
the conditions which fix the lines of constant physics.}
\begin{equation}
	\label{eq:FitFormg2SF}
	{1\over \bar{g}^2_{\rm SF}} = {1\over g_0^2} + \sum_{k=0}^{n_p}
	c_k g_0^{2k}\,.
\end{equation}
For $L_0/a=4$ we have fitted 16 data points in the ranges $\bar{g}^2_{\rm SF}=2.0451$--$1.1077$ and  
$\beta=5.9949$--$8.3130$ with $n_p=3$ obtaining $\chi^2/{\rm dof}\approx 1$.
The results for the interpolated $\beta$-values are reported in Table~\ref{tab:INparamsWS}.
For $L_0/a=6,8$, the values of $\beta$ reported in Table~\ref{tab:INparamsWS} are taken
from Table~6 of Ref.~\cite{Campos:2018ahf}, which were obtained by interpolating
the data of Ref.~\cite{DallaBrida:2018rfy} as well. As an independent check, we performed our own
fits using the functional form in Eq.~(\ref{eq:FitFormg2SF}) with $n_p=2$ including always all available data.
We obtained $\chi^2/{\rm dof}=0.38$ and $0.74$ for $L_0/a=6,8$ respectively. For the interpolated
$\beta$-values we find excellent agreement within errors between the determinations of
Ref.~\cite{Campos:2018ahf} and our results. We decided to
take as central values the results of this reference as this will allow us in the future to directly
profit from the determination of renormalization factors obtained on the ensembles
generated in Ref.~\cite{Campos:2018ahf}.
The 6 data points for $L_0/a=10$ have been fitted to the functional form in
Eq.~(\ref{eq:FitFormg2SF}) with $n_p=2$ obtaining $\chi^2/{\rm dof}\approx0.7$.
The interpolated $\beta$-values are again reported in Table~\ref{tab:INparamsWS}.
Since $T_0=2\, T_1$, the $\beta$-value of $L_0/a=4$ is the one for $T_1$ at $L_0/a=8$, while
the $\beta$-value of $L_0/a=6$ corresponds to the one for $L_0/a=12$ at $\bar g^2_{\rm SF}=1.11$
from Table 6 in Ref.~\cite{Campos:2018ahf}.
\begin{table}[t!]
  \centering
\begin{tabular}{|c|c|c|c|c|}
  \hline
  \vspace{-.4cm} &&&& \\
  $T$ & $L_0/a$ & $\beta$ &  $\kappa_{\rm cr}^{(W)}$ & $c_{\rm sw}^{(W)}$\\
\hline
\multirow{2}{*} {$T_0$} 
  &  4 & 8.7325 & 0.131887597685602 & 1.224666388699756  \\
  &  6 & 8.9950 & 0.131885781718599 & 1.214293680665697  \\
\hline                                        
\multirow{4}{*} {$T_1$} 
  &  4 & 8.3033 & 0.132316223701646 & 1.244443949720750  \\
  &  6 & 8.5403 & 0.132336064110711 & 1.233045285565058  \\
  &  8 & 8.7325 & 0.132133744093735 & 1.224666388699756  \\
  & 10 & 8.8727 & 0.131984877002653 & 1.218983546266290  \\
\hline                                        
\multirow{4}{*} {$T_2$} 
  &  4 & 7.9794 & 0.132672230374640 & 1.262303345977765  \\
  &  6 & 8.2170 & 0.132690343212428 & 1.248924515099129  \\
  &  8 & 8.4044 & 0.132476707113024 & 1.239426196162344  \\
  & 10 & 8.5534 & 0.132305706323476 & 1.232451001338001  \\
\hline                                        
\multirow{4}{*} {$T_3$} 
  &  4 & 7.6713 & 0.133039441274476 & 1.282333503658225  \\
  &  6 & 7.9091 & 0.133057201010874 & 1.266585617959733  \\
  &  8 & 8.0929 & 0.132831173856378 & 1.255711356539447  \\
  & 10 & 8.2485 & 0.132638399517155 & 1.247267216254281  \\
\hline                                        
\multirow{4}{*} {$T_4$} 
  &  4 & 7.3534 & 0.133449711446233 & 1.307002958449583  \\
  &  6 & 7.5909 & 0.133469338865844 & 1.288146969458134  \\
  &  8 & 7.7723 & 0.133228362183550 & 1.275393611340024  \\
  & 10 & 7.9322 & 0.133013578229002 & 1.265160978064686  \\
\hline                                        
\multirow{4}{*} {$T_5$} 
  &  4 & 7.0250 & 0.133908723921720 & 1.338089264736139  \\
  &  6 & 7.2618 & 0.133933679858703 & 1.315030958783770  \\
  &  8 & 7.4424 & 0.133674531074371 & 1.299622821237046  \\
  & 10 & 7.6042 & 0.133438165920285 & 1.287166774665371  \\
\hline                                        
\multirow{4}{*} {$T_6$} 
  &  4 & 6.7079 & 0.134386271436463 & 1.375352693193284  \\
  &  6 & 6.9433 & 0.134421953633166 & 1.346919223092444  \\
  &  8 & 7.1254 & 0.134141768774467 & 1.327878356622864  \\
  & 10 & 7.2855 & 0.133888442235086 & 1.312909828079458  \\
\hline                                        
\multirow{4}{*} {$T_7$} 
  &  4 & 6.3719 & 0.134926677491050 & 1.425561566301377  \\
  &  6 & 6.6050 & 0.134982857878749 & 1.389385004928746  \\
  &  8 & 6.7915 & 0.134676613758678 & 1.364706438701718  \\
  & 10 & 6.9453 & 0.134412950133538 & 1.346697162567041  \\
\hline                                        
\multirow{4}{*} {$T_8$} 
  &  4 & 6.0433 & 0.135481632961481 & 1.489790983990814  \\
  &  6 & 6.2735 & 0.135571353236717 & 1.442967721668930  \\
  &  8 & 6.4680 & 0.135236172024848 & 1.409845308468962  \\
  & 10 & 6.6096 & 0.134976206524104 & 1.388734449325687  \\
\hline
\end{tabular}
\caption{Parameters of the Monte Carlo simulations performed with the Wilson plaquette action. The bare gauge coupling is expressed in terms
  of $\beta=6/g_0^2$.}\label{tab:INparamsWS}
\end{table}

Once defined the lines of constant physics, the values of the critical mass
have been determined from Ref.~\cite{Korzec:2018}. They fix $m_{\rm cr}$ by requiring that
the PCAC mass, computed in a finite volume with SF  boundary conditions, vanishes,
see Ref.~\cite{Korzec:2018} for more details. They get
\begin{gather}
	\nonumber
  	\label{eq:mcrit_A}
 	a\mcr(g_0^2,a/L_0) = 
 	a\mcr[\rm 2lp](g_0^2,a/L_0) + c_1^{L/a} g_0^6 + c_2^{L/a} g_0^8 + c_3^{L/a} g_0^{10}\,,  
\end{gather}
were the coefficients $c_i^{L/a}$, $i=1,2,3$, are given in Ref.~\cite{Korzec:2018}.
The rest of the expression corresponds to the two-loop critical mass,
\begin{align}
	\label{eq:amcr2lpPT}
	a\mcr[\rm 2lp](g_0^2,a/L_0) &= \big(a\mcr[(0)] + \delta a\mcr[(0)](a/L_0)\big) + 
	\big( a\mcr[(1)] +\delta a\mcr[(1)](a/L_0) \big)\,g_0^2 + a\mcr[(2)] \,g_0^4  \,,
\end{align}
where
\begin{equation}\label{eq:amcrPT_plaq}
	a\mcr[(0)]=0 \;,
	\quad
	a\mcr[(1)] =  -0.270075349459 \;,
	\quad
	a\mcr[(2)] =  -0.039772       \;,
\end{equation}
are the asymptotic coefficients in the limit $L_0/a\rightarrow\infty$ while Table~\ref{tab:damcrSF} 
contains the coefficients due to cutoff effects.
The interpolated values for $\kappa_{\rm cr} =2\,am_{\rm cr} + 8$ as well as those for $c_{\rm sw}$ obtained from
Eq.~(\ref{eq:amcr2lpPT}) and Eq.~(\ref{eq:cSW-W}) respectively are reported in
Table~\ref{tab:INparamsWS} and are indicated with $\kappa_{\rm cr}^{(W)}$ and $c_{\rm sw}^{(W)}$.

\subsection{Low temperatures}
\begin{table}[t!]
	\centering
	\small
	\begin{tabular}{|c|c|c|}
		\hline
		$T$ &  $\bar{g}^2_{\rm GF}(\mu=T/\sqrt{2})$ & $T$ (GeV) \\
		\hline
		$T_9$     & 2.7359  & 2.833(68) \\    
		$T_{10}$  & 3.2029  & 1.821(39) \\   
		$T_{11}$  & 3.8643  & 1.167(23) \\   
		\hline
	\end{tabular}
	\caption{Values of the GF couplings corresponding to the 
		lines of constant physical temperature that we consider.}
	\label{tab:RenCouplingValuesGF}	
\end{table}
The lower temperature values $T_9, T_{10}$ and $T_{11}$ are fixed analogously to
the higher ones but from the gradient flow (GF) coupling. The
temperature is fixed by imposing that 
\begin{equation}
	\label{eq:T2MuGF}
	T={1\over L_0\sqrt{2}}=\sqrt{2} \mu\,,
\end{equation}
where $\mu$ is the renormalization scale of the GF coupling $\bar{g}^2_{\rm GF}(\mu)$
defined in a box with spatial and temporal extensions satisfying
$L^{\rm GF} = L^{\rm GF}_0  = 1/\mu$, i.e. $L_0=L^{\rm GF}_0/2$.

In order to determine the physical values of the temperature, we start from
the result (cf.~Eqs.~(15)-(16) and Tables I-II of
Ref.~\cite{Bruno:2017gxd}),
\begin{equation}
	\label{eq:muhad}
	\bar{g}^2_{\rm GF}(\mu_{\rm had, 1})=11.31
	\quad
	\Rightarrow
	\quad
	\mu_{\rm had, 1}=196.9(3.2)\,{\rm MeV}\,,
\end{equation}
where $\mu_{\rm had, 1}$ is 
inferred from the experimental value of a combination of the pion and
kaon decay constant as for $\mu_0$. The value of the temperatures corresponding to the
couplings of interest can then be inferred through the relation,
\begin{equation}
	\label{eq:MuRatioGF}
	\ln\bigg({\mu\over\mu_{\rm had, 1}}\bigg)=\int_{\bar{g}_{\rm GF}(\mu_{\rm had, 1})}^{\bar{g}_{\rm GF}(\mu)}{{\rm d }g\over \beta_{\rm GF}(g)}\,,
\end{equation}
where
\begin{equation}
	\mu{{\rm d}\bar{g}_{\rm GF}(\mu)\over{\rm d}\mu}=\beta_{\rm GF}(\bar{g}_{\rm GF})\,.
\end{equation}
Using the results of Ref.~\cite{DallaBrida:2016kgh}, the non-perturbative
$\beta$-function of the GF coupling can be parameterized over the range 
of couplings of interest as (cf.~Eq.~(2.36) of Ref.~\cite{Campos:2018ahf})
\begin{equation}
	\beta_{\rm\scriptscriptstyle GF}(\bar{g}) = -\frac{\bar{g}^3}{\sum_{n=0}^2 p_n\bar{g}^{2n}} \,, 
	\qquad \bar{g}^2 \in [2.1,11.3]\,,
\end{equation}
with fit parameters
\begin{equation}
	\label{eq:gfparm}
	p_0 = 16.07\,,\quad p_1 = 0.21\,,\quad p_2=-0.013\,,
\end{equation}
and covariance matrix 
\begin{equation}
	\label{eq:gfcov}
	{\rm cov}(p_i,p_j) = \left(
	\begin{array}{rrr}
		5.12310\times 10^{-1}& -1.77401\times 10^{-1}& 1.32026\times 10^{-2}\\
		-1.77401\times 10^{-1}& 6.60392\times 10^{-2}& -5.10305\times 10^{-3}\\
		1.32026\times 10^{-2}& -5.10305\times 10^{-3}& 4.06114\times 10^{-4}\\
	\end{array}
	\right)\,.
\end{equation}
Given this representation, we integrated Eq.~(\ref{eq:MuRatioGF}) numerically
using the result for $\mu_{\rm had, 1}$ in Eq.~(\ref{eq:muhad}), the relation
Eq.~(\ref{eq:T2MuGF}), and the values for the coupling in Table~\ref{tab:RenCouplingValuesGF}
where are also reported the final values of the temperatures $T_9, T_{10}$ and $T_{11}$.

\subsubsection{Bare parameters}
For the 3 lowest temperatures $T_9$, $T_{10}$ and $T_{11}$ we adopted the 
tree-level Symanzik improved gauge action in Eq.~(\ref{eq:cSW-I}) so as
to be able to use the results from Ref.~\cite{DallaBrida:2016kgh} on
the GF coupling, $\bar{g}^2_{\rm GF}(\mu)$, computed in the massless theory.

For each value of $L_0/a$,  the bare parameters are taken from Table~8 of
Ref.~\cite{Campos:2018ahf} and are reported in Table~\ref{tab:INparamsGF}.
To verify that the temperature is constant within
each set, we have fitted the results in Table~1 of
Ref.~\cite{DallaBrida:2016kgh} for each value of $L_0/a$ using the functional form
in Eq.~(\ref{eq:FitFormg2SF}) but
with $\bar{g}^2_{\rm SF}(\mu)$ replaced by
$\bar{g}^2_{\rm GF}(\mu)$ and by taking into account that in this case $\mu=1/(2 L_0)$. By including
all the 9 data points for each value of $L_0/a$, and by choosing $n_p=2$ for $L_0/a=4,6$, and $n_p=3$ for $L_0/a=8$,
we obtained excellent fits with $\chi^2/{\rm dof}\approx 0.93$, $0.16$ and $1.07$ for
$L_0/a=4$, $6$ and $8$ respectively. The results confirm that the temperature is constant
within errors for the lattices within each set\footnote{Considerations analogous to those in
footnote 3 apply also here for the case of the GF coupling.}.

Once the lines of constant physics have been defined, the corresponding values of the critical mass
have been computed from the result in Appendix A.1.4 of
Ref.~\cite{DallaBrida:2016kgh} which reads
\begin{align}
	\label{eq:amcresult}
	a\mcr(g^2_0,a/L_0) &= \left( {\sum}_{k=0}^{6} \mu_k\, g_0^{2k} \right) \times \left( {\sum}_{i=0}^{6} \zeta_i \, g_0^{2i} \right)^{-1} \,,
\end{align}
with the parameters $\mu_k$ and $\zeta_i$ listed in Table
\ref{tab:fitparams} for the relevant $L_0/a$. As for the case of the 
Wilson-plaquette gauge action, the values of $m_{\rm cr}(g^2_0,a/L_0)$ 
depend on $L_0/a$ because it has been determined by requesting the PCAC 
mass to vanish in a finite volume, see Ref.~\cite{DallaBrida:2016kgh} for more details.

\begin{table}[t!]
	\centering
	\small
	\begin{tabular}{|c|c|c|c|} \hline
		Coeff.     & $L_0/a=4$              & $L_0/a=6$             & $L_0/a=8$             \\ \hline
		$\zeta_0$  & $+1.005834130000000$ & $+1.002599440000000$ & $+1.001463290000000$ \\
		$\mu_0$    & $-0.000022208694999$ & $-0.000004812471537$ & $-0.000001281872601$ \\
		$\mu_1$    & $-0.202388398516844$ & $-0.201746020772477$ & $-0.201520105247962$ \\ \hline
		$\zeta_1$  & $-0.560665657872021$ & $-0.802266237327923$ & $-0.892637061391273$ \\
		$\zeta_2$  & $+3.262872842957498$ & $+4.027758778155415$ & $+5.095631719496583$ \\
		$\zeta_3$  & $-5.788275397637978$ & $-6.928207214808553$ & $-8.939546687871335$ \\
		$\zeta_4$  & $+4.587959856400246$ & $+5.510985771180077$ & $+7.046607832794273$ \\
		$\zeta_5$  & $-1.653344785588201$ & $-2.076308895962694$ & $-2.625638312722623$ \\
		$\zeta_6$  & $+0.227536321065082$ & $+0.320430672213824$ & $+0.405387660384441$ \\
		$\mu_2$    & $+0.090366980657738$ & $+0.128161834555849$ & $+0.139461345465939$ \\
		$\mu_3$    & $-0.600952105402754$ & $-0.681097059845447$ & $-0.847457204378732$ \\
		$\mu_4$    & $+0.934252532135398$ & $+0.991316994385556$ & $+1.261676178806362$ \\
		$\mu_5$    & $-0.608706158693056$ & $-0.606597739050552$ & $-0.754644691612547$ \\
		$\mu_6$    & $+0.140501978953879$ & $+0.129031928169091$ & $+0.153135714480269$ \\
		\hline
	\end{tabular}
	\caption{Coefficients for the parameterization	Eq.~\eqref{eq:amcresult}. 
		The three leading coefficients $\zeta_0$, $\mu_0$, and $\mu_1$
                in the upper part of the table are combinations of known perturbative
                coefficients while the  others were determined by a fit.}
	\label{tab:fitparams}
\end{table}

Once the $\beta$-values have been determined, the corresponding values for
$\kappa_{\rm cr} =2\,am_{\rm cr} + 8$ as well as those for $c_{\rm sw}$ are obtained from
Eqs.~(\ref{eq:amcresult}) and (\ref{eq:cSW-I}) respectively, and are reported in
Table~\ref{tab:INparamsGF} as $\kappa_{\rm cr}^{(I)}$ and $c_{\rm sw}^{(I)}$.
\begin{table}[t!]
  \centering
\begin{tabular}{|c|c|c|c|c|}
\hline
\vspace{-.4cm} &&&& \\
 $T$  & $L_0/a$ & $\beta$ &  $\kappa_{\rm cr}^{(I)}$ & $c_{\rm sw}^{(I)}$\\
\hline
\multirow{3}{*} {$T_9$} 
  &  4 & 4.764900 & 0.134885548000448 & 1.335350323996506 \\
  &  6 & 4.938726 & 0.134507608658235 & 1.308983384364439 \\
  &  8 & 5.100000 & 0.134168886219319 & 1.288203306487197 \\
\hline
\multirow{3}{*} {$T_{10}$} 
  &  4 & 4.457600 & 0.135606746160064 & 1.39574103127591 \\
  &  6 & 4.634654 & 0.135199857298424 & 1.358462476494125 \\
  &  8 & 4.800000 & 0.134821158536685 & 1.329646151978636 \\
\hline
\multirow{3}{*} {$T_{11}$} 
  &  4 & 4.151900 & 0.136325892438363 & 1.482418125298923 \\
  &  6 & 4.331660 & 0.135926636004668 & 1.427424655158656 \\
  &  8 & 4.500000 & 0.135525721037715 & 1.386110343557152  \\
\hline
\end{tabular}
\caption{Parameters of the Monte Carlo simulations performed with the tree-level improved Symanzik action.
The bare gauge coupling is  expressed in terms of $\beta=6/g_0^2$.}\label{tab:INparamsGF} 
\end{table}

\section{Finite-volume effects in thermal two-point correlators\label{app:finiteV}}
In this appendix we derive the formula for the leading finite-volume effects
in the spatial correlators $C_\Obs(x_3)$ defined in Eq.~(\ref{eq:2pt}) at asymptotically high temperatures.
We follow the lines of argumentation in Refs.~\cite{Meyer:2009kn,Giusti:2012yj,Hansen:2019rbh},
and we assume the reader to be familiar with these papers. Gauge and
quark fields are assumed to satisfy shifted boundary conditions, Eqs.~(\ref{eq:shift_gluons}) and (\ref{eq:shift_quark}),
with $\boldsymbol{\xi}=(\xi_1,0,0)$. In the continuum theory, where finite volume effects
are derived, the results can be readily generalized to a generic shift $\boldsymbol{\xi}$
by exploiting the invariance of the theory under the $SO(3)$ spatial rotations, see
Ref.~\cite{Giusti:2012yj} for details. 

We start by considering a box of volume $L_0 \times L^3$, and we define the finite-volume
residue due to the compactification in the $1$-direction as 
\begin{equation}
	\label{eq:I1}
	\mathcal{I}_1(x_3,L)\equiv 
	\Big[1-\lim_{L_1\to\infty}\Big]C_\Obs(x_3)\,,
\end{equation}
where $L_1$ is the length of the box in direction $1$. In order to determine $\mathcal{I}_1$
we consider the transfer-matrix representation  of $C_\Obs(x_3)$ along the $1$-direction
(cf.~Sect.~$4$ of Ref.~\cite{Giusti:2012yj}),
\begin{equation}
	\label{eq:Cv1}
	C_\Obs(x_3)=
	\int dx_0  dx_1 dx_2\,
	{\Tr[e^{-(L\gamma_1-x_1)\widetilde{H}}\Obs^a(\tilde{\boldsymbol{x}})\,
	e^{-x_1\widetilde{H}}\Obs^a(\tilde{\boldsymbol{0}})\,
	e^{-iL\gamma_1\xi_1\widetilde{\omega}}]
	\over \Tr[e^{-L\gamma_1(\widetilde{H}+i\xi_1\widetilde{\omega})}]}\,,
\end{equation}
where $\tilde{\boldsymbol{x}}=(x_0,x_2,x_3)$, $\gamma_1=(1+\xi_1^2)^{-1/2}$, and the trace $\Tr$ is carried over the states of
the corresponding Hilbert space. In this equation,
$\widetilde{H}$ stands for the screening Hamiltonian along 
the $1$-direction. This operator has a discrete spectrum of states defined on a 
slice of dimensions $(L_0/\gamma_1) \times L\times L $, with ordinary periodic 
boundary conditions. The operator $\widetilde{\omega}$ denotes instead the momentum operator 
along the 0-direction of length $(L_0/\gamma_1)$. We indicate with $|n\rangle$ 
the simultaneous eigenstates of $\widetilde{H}$, $\widetilde{\omega}$, and potentially 
other conserved charge operators. The eigenvalues of $\widetilde{H}$ and 
$\widetilde{\omega}$ corresponding to the state $|n\rangle$ are the energies 
$E_n$ and Matsubara frequencies ${\omega}_n=2\pi m_n\gamma_1/L_0$, $m_n\in\mathbb{Z}$, 
respectively. We assume that the states are ordered in such a way that $E_{n+1}\geq E_n$. 
The state $|0\rangle$ is therefore the unique ground state of the system, for which 
we conveniently set $E_0=0$. At asymptotically high temperature, the state $|1\rangle$ is then 
expected to have a strictly positive mass(-gap), $M_{\rm gap}$, proportional to the
temperature $T$, see section~\ref{sec:stgy1}. 
Furthermore, due to the fact that at asymptotically high temperature the effective theory 
of QCD contains only purely gluonic degrees of freedom, we expect the lowest-lying 
energy states, i.e.~those with $E_n\ll \pi T$, to have zero flavour
quantum-numbers~\cite{Laine:2009dh}. Inserting two complete sets of eigenstates $|n\rangle$
with zero baryon number in Eq.~(\ref{eq:Cv1}), we have
\begin{equation}
	\label{eq:Cv2}
        C_\Obs(x_3)\!=\!{1\over Z}\!\!
	\int\!\! dx_0dx_1dx_2\!
	\sum_{n,n'}e^{-L\gamma_1(E_n+i\xi_1\omega_{n})} 
	e^{-x_1(E_{n'}-E_n)}
	\langle n | \Obs^a(\tilde{\boldsymbol{x}})|n'\rangle 
	\langle n'|\Obs^a(\tilde{\boldsymbol{0}})|n\rangle + \dots \,,
\end{equation}
where $Z=\sum_{n} e^{-L\gamma_1(E_{n}+i\xi_1\omega_{n})}+\dots$, and the dots stand for baryonic contributions
which are suppressed exponentially with respect to the sum.
Let us focus on the terms in the sum for which $E_n\neq E_{n'}$. For these, the integral over $x_1$ gives
\begin{equation}
	e^{-L\gamma_1E_n}\int_{0}^{L\gamma_1} dx_1\, e^{-x_1(E_{n'}-E_n)}=
	{e^{-L\gamma_1E_n}-e^{-L\gamma_1E_{n'}}
	\over E_{n'}-E_n}\,.
\end{equation}
Inserting this relation in Eq.~(\ref{eq:Cv2}), and relabeling $n\leftrightarrow n'$
in some terms, we obtain
\begin{multline}
	\label{eq:Cv3}
	C_\Obs(x_3)={1\over Z}
	\int dx_0 dx_2\,
	\sum_{\substack{n,n'\\E_n\neq E_{n'}}} 
	{e^{-L\gamma_1(E_n+i\xi_1\omega_{n})}\over E_{n'}-E_n}\times\\
	\times
	\Big\{\
	\langle n | \Obs^a(\tilde{\boldsymbol{x}})|n'\rangle
	\langle n'|\Obs^a(\tilde{\boldsymbol{0}})|n\rangle + 
	\langle n | \Obs^a(\tilde{\boldsymbol{0}})|n'\rangle
	\langle n'|\Obs^a(\tilde{\boldsymbol{x}})|n\rangle 
	\Big\}+ \ldots ,
\end{multline}
where the terms with $E_n=E_{n'}$ are included in the dots.%
\footnote{Note that in order to derive Eq.~(\ref{eq:Cv3}) we used the fact that $C_\Obs(x_3)$ 
is projected onto zero Matsubara frequency and therefore $\omega_{n}=\omega_{n'}$.} 
In this form, it is evident that in the limit where $L_1\to\infty$, 
the terms with energies $E_n\ll \pi T$ dominate the sum. Furthermore, within this energy 
range, there are no terms with $E_n=E_{n'}$ that can contribute. This is because the 
operators $\Obs^a$ have non-trivial flavour quantum-numbers and any flavoured mesonic state has 
an energy $E_n\gtrsim 2\pi T$. As we are interested in determining the leading finite-volume 
effects in $C_\Obs(x_3)$, from now on we shall restrict ourselves to consider only states 
that satisfy the above energy constraint. These include, in particular, the theory vacuum 
and the 1-particle states with mass equal to the mass-gap $M_{\rm gap}$. 
We can thus introduce the two-point correlation function
\begin{equation}
	G_n(\tau,\tilde{\boldsymbol{x}})=
	\langle n |{\rm T}\{ \Obs^a(\tau,\tilde{\boldsymbol{x}})\Obs^a(0,\tilde{\boldsymbol{0}})\}|n\rangle\,,
\end{equation}
where ${\rm T}\{\cdots\}$ stands for the ordered product of the operators with respect 
to the parameter $\tau$, and $\Obs^a(\tau,\tilde{\boldsymbol{x}})=
e^{\tau\widetilde{H}}\,\Obs^a(\tilde{\boldsymbol{x}})\,e^{-\tau\widetilde{H}}$.
After some trivial algebra, it is immediate to show that,
\begin{equation}
	\int_{-\infty}^\infty\!\!\!\!\! d\tau\,G_n(\tau,\tilde{\boldsymbol{x}})\!=\!
	\sum_{n'}{1\over E_{n'}-E_n}
	\Big\{
	\langle n | \Obs^a(\tilde{\boldsymbol{x}})|n'\rangle
	\langle n'|\Obs^a(\tilde{\boldsymbol{0}})|n\rangle + 
	\langle n | \Obs^a(\tilde{\boldsymbol{0}})|n'\rangle
	\langle n'|\Obs^a(\tilde{\boldsymbol{x}})|n\rangle 
	\Big\}\,.
\end{equation}
Considering the $L_1\to\infty$ limit of Eq.~(\ref{eq:Cv3}), and using the 
above relation, we find for $\mathcal{I}_1$ the result
\begin{equation}
	\label{eq:I1Final}
	\mathcal{I}_1(x_3,L)=
	\sum_{n\big|\substack{\text{1-particle}\\\text{states}}}
	e^{-L\gamma_1(E_n+i\xi_1\omega_{n})}
	\int dx_0 dx_2 d\tau\,\big\{G_n(\tau,\tilde{\boldsymbol{x}})-G_0(\tau,\tilde{\boldsymbol{x}})\big\}
	+\ldots\,,
\end{equation}
where the energies of the 1-particle states are confined to the range $M_{\rm gap}\lesssim E_n \lesssim \pi T$ 
and the dots stand for terms which are exponentially suppressed compared to the leading ones.
From this expression, it is immediate to conclude that $\mathcal{I}_1$ is exponentially 
suppressed as $M_{\rm gap}L\to\infty$. Moreover the length of the other two spatial directions can be sent
to infinity on the r.h.s. of Eq.~(\ref{eq:I1Final}) up to sub-leading finite-volume effects. The analogous
contribution from the 2-direction, $\mathcal{I}_2$, is obtained from the one for $\mathcal{I}_1$ in
Eq.~(\ref{eq:I1Final}) by replacing $L\gamma_1\to L$, $\xi_1\to0$, and $x_2\to x_1$. (Note that the length of
the 0-direction remains $L_0/\gamma_1$). Finite-volume corrections in the 3-direction can be taken into account,
as usual, by considering the backward propagation in the series of exponentials due to periodic
boundary conditions in that direction. The total finite-volume effects in $C_\Obs(x_3)$ are finally given
by summing all three contributions. Since the screening masses are extracted at asymptotically large distances
according to Eq.~(\ref{eq:mscrn}), their finite-volume corrections are determined by $\mathcal{I}_1+\mathcal{I}_2$
only.

\section{Inversion of the Dirac operator\label{app:Dinv}}
The usual stopping criterion used in iterative methods for the numerical solution of the Dirac equation
\be
D\psi=\eta
\ee
requires that the norm of the residual $\rho = D\psi - \eta$ is sufficiently small, i.e. the global condition
$r=\norm{\rho}/\norm{\eta}<\varepsilon$.  The tolerance $\varepsilon$  is chosen to be small enough that the
error introduced by using such an approximate solution must be negligible with respect to the statistical fluctuations
on the observable of interest.  However, the tolerance cannot be smaller than what is allowed by the
finite-precision arithmetic of a given implementation.

At high temperature, the lowest Matsubara frequency $\pi T$ provides an infrared cutoff to quark propagation. As a
result, the matrix elements $D^{-1}(x,y)$ become extremely small when $T\, |x-y| \gg 1$, and a very accurate solution
of the Dirac equation is required at those distances. The brute-force approach of simply implementing higher-precision
and requiring a smaller tolerance is not always practicable. In this case, a solution is achieved by introducing
a preconditioned version of the Dirac equation,
\be
\tilde D \tilde \psi=\tilde \eta
\ee
where
\be\label{eq:dp}
\tilde D = M^{-1} D M\;, \qquad \tilde \psi= M^{-1}\psi\;, \qquad  \tilde \eta= M^{-1}\eta\; ,
\ee
with the preconditioning matrix $M$ chosen so that the various components of the solution $\tilde\psi$ are comparable
in magnitude~\cite{deDivitiis:2010ya}.  

The quark propagators needed for the two-point meson correlation functions considered in this paper have been computed
by implementing the preconditioning matrix 
\begin{align}
    M(x,y) &= \cosh\{m_M (x_3-y_3-L/2)\} \cdot \id,
    \label{eq:dp_mz}
\end{align}
where $\id$ indicates the identity matrix in the indices not explicitly indicated, i.e. color, spin and
the first three components of space-time coordinates. After some tuning, for the lattices with $L_0/a=4, 6, 8$ and $10$
we have chosen $m_M=0.4, 0.3, 0.2$ and $0.15$ respectively, with the shift being always $\bsxi=(1,0,0)$. This indeed
guarantees that the components of $\tilde\psi$ are always comparable in magnitude. We have also monitored explicitly
a posteriori that the global condition $r<\varepsilon$ is always satisfied by the solution vector.

\section{Simulation details and results\label{app:OUT}}
We have simulated three-flavour QCD with a HMC algorithm by using the \texttt{openQCD-1.6}
package~\cite{Luscher:2012av,openQCD1.6} modified so as to allow for shifted boundary conditions.
\begin{table}[th!]
\centering
\begin{tabular}{|c|c|c|c|c|l|l|l|}
\hline
&  &  &  &  &  &  & \\[-0.125cm]
  $T$ & $L_0/a$ & $n_{\rm mdu}$ & $n_{\rm skip}$ & $n_{\rm nsrc}$   &$\;\;\;\displaystyle\frac{m_P}{2\pi T}$\!\!\! &
  $\;\;\;\displaystyle \frac{m_V}{2\pi T}$\!\!\! & $\!\!\displaystyle\frac{(m_V\! - \!m_P)}{2\pi T}\!\!$\\[-0.125cm]
&  &  &  &  &  &  & \\
\hline
\multirow{2}{*} {$T_0$} 
  &  4 & 90 & 10 & 4 & 0.9659(5) &  0.9716(7) & 0.00577(20) \\
  &  6 & 90 & 10 & 2 & 0.9934(14) & 0.9996(12)& 0.0065(4) \\
\hline                                        
\multirow{4}{*} {$T_1$} 
  &  4 & 90 & 10 & 4 & 0.9656(7) &  0.9721(8)& 0.0068(3) \\
  &  6 & 270 & 30 & 2 & 0.9945(14) & 1.0014(19)& 0.0070(8) \\
  &  8 & 450 & 50 & 2 & 1.0078(18) & 1.0148(20)& 0.0075(5) \\
  & 10 & 900 & 100 & 2 & 1.0090(25) & 1.0160(27)& 0.0075(4) \\
\hline                                        
\multirow{4}{*} {$T_2$} 
  &  4 & 90 & 10 & 4 & 0.9685(7) &  0.9753(8)& 0.0075(3) \\
  &  6 & 270 & 30 & 2 & 0.9961(14) & 1.0049(18)& 0.0089(5) \\
  &  8 & 450 & 50 & 2 & 1.0055(23) & 1.0147(25)& 0.0089(5) \\
  & 10 & 900 & 100 & 2 & 1.0122(25) & 1.0207(25)& 0.0073(6) \\
\hline                                        
\multirow{4}{*} {$T_3$} 
  &  4 & 90 & 10 & 4 & 0.9682(11) & 0.9764(18)& 0.0087(5) \\
  &  6 & 270 & 30 & 2 & 0.9971(11) & 1.0050(16)& 0.0084(10) \\
  &  8 & 450 & 50 & 2 & 1.0039(18) & 1.0130(22)& 0.0083(7) \\
  & 10 & 810 & 90 & 2 & 1.0124(25) & 1.0219(29)& 0.0099(7) \\
\hline                                        
\multirow{4}{*} {$T_4$} 
  &  4 & 90 & 10 & 4 & 0.9704(7) & 0.9804(14)& 0.0103(4) \\
  &  6 & 270 & 30 & 2 & 0.9973(14) & 1.0087(14)& 0.0109(8) \\
  &  8 & 450 & 50 & 2 & 1.0051(20) & 1.0172(25)& 0.0093(9) \\
  & 10 & 540 & 60 & 2 & 1.0138(20) & 1.0248(23)& 0.0108(7) \\
\hline                                        
\multirow{4}{*} {$T_5$} 
  &  4 & 90 & 10 & 4 & 0.9708(8) &  0.9838(12)& 0.0128(4)\\
  &  6 & 180 & 20 & 2 & 0.9941(22) & 1.006(3)& 0.0109(20) \\
  &  8 & 450 & 50 & 2 & 1.0057(18) & 1.0172(29)& 0.0119(21) \\
  & 10 & 540 & 60 & 2 & 1.0090(27) & 1.0228(29)& 0.0137(10) \\
\hline                                        
\multirow{4}{*} {$T_6$} 
  &  4 & 90 & 10 & 4 & 0.9676(10) & 0.9830(18)& 0.0156(11) \\
  &  6 & 180 & 20 & 2 & 0.9948(15) & 1.0089(24)& 0.0142(11) \\
  &  8 & 450 & 50 & 2 & 1.0037(29) & 1.018(4)& 0.0150(23) \\
  & 10 & 540 & 60 & 2 & 1.0108(25) & 1.026(4)& 0.0153(16) \\
\hline                                        
\multirow{4}{*} {$T_7$} 
  &  4 & 90 & 10 & 4 & 0.9679(8) & 0.9854(18)& 0.0172(11) \\
  &  6 & 180 & 20 & 2 & 0.9930(15) & 1.0093(28)& 0.0171(17) \\
  &  8 & 450 & 50 & 2 & 1.0051(22) & 1.024(4)& 0.0188(16) \\
  & 10 & 900 & 100 & 2 & 1.012(3) & 1.028(5)& 0.0171(19) \\
\hline                                        
\multirow{4}{*} {$T_8$} 
  &  4 & 90 & 10 & 4 & 0.9677(8) &  0.9910(18)& 0.0235(17) \\
  &  6 & 180 & 20 & 4 & 0.9907(16) & 1.015(4)& 0.0237(17) \\
  &  8 & 450 & 50 & 4 & 1.000(3) & 1.025(4)& 0.0247(14) \\
  & 10 & 900 & 100 & 4 & 1.0032(23) & 1.0288(25)& 0.0252(14) \\
\hline
\end{tabular}
\hfill
\caption{Results for the pseudoscalar, $m_P$, and the vector, $m_V$, non-singlet meson masses together with
their difference $(m_V\! - \!m_P)$ all normalized to $2\pi T$ at finite lattice spacing for the
temperatures $T_0, \ldots, T_8$.
The number of MDUs generated, $n_{\rm mdu}$, those
skipped between two consecutive measurements, $n_{\rm skip}$, and the number of local sources per
configuration on which the two-point correlation functions have been computed, $n_{\rm nsrc}$, are also
reported.}\label{tab:OUTmassesH}
\end{table}
We have employed several efficient algorithms to speed up the simulations. More precisely,
the doublet of up and down quarks have been simulated with an optimized twisted-mass Hasenbusch
preconditioning of the quark determinant~\cite{Hasenbusch:2001ne,Luscher:2012av}. The determinant
has been split in three factors by employing the twisted masses values $a\mu=0.0, 0.1$ and $1.0$.
The strange quark has been simulated through a RHMC algorithm~\cite{Kennedy:1998cu,Clark:2003na} with an
optimized frequency splitting of the rational approximation in two separate contributions. Even-odd 
preconditioning has been used for both the light and strange quarks. The integration of the molecular dynamics
equations has been based on a three-level integration scheme. The gauge force has been integrated on the finest level
using a 4th-order Omelyan-Mryglod-Folk (OMF4) integrator~\cite{Omelyan:2013} with step-size 1, while the
fermionic forces have been integrated on the two coarser levels. On the finest of these we have used a OMF4
integrator step-size 1, while  on the coarsest a 2nd-order OMF
integrator~\cite{Omelyan:2013} with step-size between 7 and 9. The solution of the Dirac equation along the molecular
dynamics evolution has been obtained by using a standard conjugate gradient with chronological inversion. The
length of each trajectory is 2 MDUs for all lattices. More details on the exact implementation of these
algorithms can be found in Refs.~\cite{Luscher:2012av,openQCD1.6}.

\begin{table}[t!]
\centering
\begin{tabular}{|c|c|c|c|c|l|l|l|}
  \hline
  &  &  &  &  &  &  & \\[-0.125cm]
  $T$ & $L_0/a$ & $n_{\rm mdu}$ & $n_{\rm skip}$ & $n_{\rm nsrc}$   &$\;\;\;\displaystyle\frac{m_P}{2\pi T}$\!\!\! &
  $\;\;\;\displaystyle \frac{m_V}{2\pi T}$\!\!\! & $\!\!\displaystyle\frac{(m_V\!-\!m_P)}{2\pi T}\!$\\[-0.125cm]
  &  &  &  &  &  &  & \\
\hline
\multirow{3}{*} {$T_9$} 
  &  4 & 90 & 10 & 4 & 0.9663(16) & 0.9872(23) & 0.0205(15) \\
  &  6 & 90 & 10 & 4 & 0.9907(24) & 1.012(4) & 0.0233(27) \\
  &  8 & 90 & 10 & 4 & 1.0010(20) & 1.0238(25) & 0.0233(16) \\
\hline
\multirow{3}{*} {$T_{10}$} 
  &  4 & 90 & 10 & 4 & 0.9645(13) & 0.9912(17) & 0.0259(22) \\
  &  6 & 90 & 10 & 4 & 0.9896(11) &  1.0203(24)& 0.0294(18) \\
  &  8 & 90 & 10 & 4 & 0.9963(22) & 1.024(4) & 0.0290(16) \\
\hline
\multirow{3}{*} {$T_{11}$} 
  &  4 & 90 & 10 & 4 & 0.9552(16) & 0.992(3)  & 0.0375(18) \\
  &  6 & 90 & 10 & 8 & 0.9768(20) & 1.018(5)  & 0.0406(26) \\
  &  8 & 90 & 10 & 8 & 0.9912(16) &  1.031(6) & 0.039(4) \\
\hline 
\end{tabular}
\hfill
\caption{As in Table~\ref{tab:OUTmassesH} but for $T_9$, $T_{10}$ and $T_{11}$.\label{tab:OUTmassesL}}
\end{table}

For each ensemble, we have started the thermalization phase by simulating a lattice with a spatial length of $L/a=48$
in all three directions and the same bare parameters as the target one. After approximately $1000$ MDUs, we have
duplicated the lattice in each direction so that $L/a=96$. We have then run the HMC for approximately $500$ MDUs, after
which we have triplicated the lattice in all spatial directions so to have $L/a=288$. We have completed the thermalization phase
by running the HMC for a number of MDUs between $100$ to $200$, and then we have started the computation
of the correlation functions. During all the phases of thermalization we have always monitored
the action and the various components of the energy-momentum tensor. We have also constantly monitored the topological charge
computed with the Wilson flow, and we have explicitly checked that at the end of each thermalization process we always ended up 
in the trivial topological sector.

Once the thermalization has been concluded, we have accumulated a certain number of configurations for the computation
of the EoS. Among those, we have selected some that we have used for the computation of the screening masses.
In particular in Tables~\ref{tab:OUTmassesH} and \ref{tab:OUTmassesL} we report the number of MDUs considered,
the number of MDUs skipped between two consecutive independent configurations, and the number of local sources per
configuration on which the two-point correlation functions have been computed. For each configuration, the best
estimates of $C_{{\cal O}}(x_3)$ in Eq.~(\ref{eq:2pt_lat}) have been obtained by properly averaging their values from
all local sources, and then symmetrizing the correlators with respect to $x_3=L/2$. The screening masses have then been
extracted as described in Section~\ref{sect:num_study}. The results are
reported in Tables~\ref{tab:OUTmassesH} and \ref{tab:OUTmassesL} for the 9 highest temperatures
$T_0, \ldots$, $T_8$ and for the lowest ones, $T_9$, $T_{10}$ and $T_{11}$ respectively. 

To explicitly check that finite volume effects are negligible within our statistical errors, we have
generated three more lattices at $T_0$ ($L_0/a=6$), $T_1$ ($L_0/a=10$) and $T_{11}$ ($L_0/a=8$) at
three smaller spatial volumes, namely $6\times 144^2\times 288$, $10\times 96^2 \times 288$, and
$8\times 144^2 \times 288$ (direction $3$ the longest) respectively. On these lattices we have computed the screening
masses following the same procedure as described before. They are in very good agreement with the analogous ones
reported in Tables ~\ref{tab:OUTmassesH} and \ref{tab:OUTmassesL}, and therefore they confirm the
theoretical expectations that finite volume effects are negligible.

\section{Screening masses in the free lattice theory\label{app:mass_SB}}
With the aim of accelerating the continuum limit extrapolation, we report here
the calculation of the non-singlet meson screening masses in the free theory on
the lattice. Since $D_{\rm sw}$ in Eq.~(\ref{eq:Dirac}) does not contribute in the
free case, the quark propagator in momentum space for each single flavour
is given by 
\begin{equation}
  S(p) =  \frac{-i \gamma_\mu \, \bar p_\mu  + {m}_0(p)}{D_F(p)}\; ,
  \qquad \mbox{with} \qquad
  D_F(p) = \sum_{\mu=0}^3\bar p_\mu^2 + m_0^2(p)
\end{equation}
and
\begin{equation}
  m_0(p) =  m_0 + \frac{a}{2} \sum_{\mu=0}^3 \hat p^2_\mu\; ,
  \qquad
  \bar p_\mu= {1 \over a}\sin( a p_\mu) \; ,
  \qquad
\hat p_\mu = { 2 \over a}
\sin\left( \frac{a p_\mu}{2}\right)\; ,
\end{equation}
where we have assumed $M_0=m_0\!\cdot\!\id$. In the
presence of shifted boundary conditions (see appendices A and E in~\cite{DallaBrida:2020gux}),
the fermionic lattice momenta in the compact direction take the values
\begin{equation}\label{eq:MOMferm1}
p_0 = \frac{2\pi n_0 }{ L_0} + \frac{\pi}{ L_0}-\sum_{k=1}^3 p_k \xi_k\; 
  \quad \mbox{where} \quad
  n_0=0,\ldots,L_0/a-1
\end{equation}
while in the spatial directions we consider the infinite volume limit and therefore the momenta are
given by $p_k \in [-\pi/a,\pi/a)$.
\begin{table}[t!]
  \centering
\begin{tabular}{|c|c|}
  \hline
  $L_0/a$ & $m^{\rm free}/(2\pi T)$\\
\hline
4  & 0.932614077\ldots \\
6  & 0.967811412\ldots \\
8  & 0.981401809\ldots \\
10 & 0.987944825\ldots \\
\hline
\end{tabular}
\caption{Tree-level values of the non-singlet screening masses on lattices with temporal extension $L_0/a$, infinite
spatial volume, and shift vector $\bsxi=(1,0,0)$.\label{tab:TreeM}}
\end{table}

To extract the screening masses, we compute
the two-point correlators defined in Eq.~(\ref{eq:2pt}) for ${\cal O}=\left\{S,P,V_\mu,A_\mu\right\}$
corresponding to $\Gamma_{{\cal O}}=\left\{\id,\gamma_5,\gamma_\mu,\gamma_\mu\gamma_5\right\}$ respectively.
At tree-level they are given by
\begin{equation}
C_{{\cal O}}(x_3) = -\frac{3}{2} \int \frac{d^4 p}{(2\pi)^4}\frac{d k_3}{2\pi}\, \Tr \Big[ \Gamma_{{\cal O}}\, S(k)\, \Gamma_{{\cal O}}\, S(p) \Big] e^{-i(p_3-k_3)\, x_3}\; ,
\end{equation}
where $k=(p_0,p_1,p_2,k_3)$ and $\Tr$ stands for the trace over the Dirac index. From Eq.~($3.613-1.^6$)
of Ref.~\cite{gradshteyn2007} we obtain  
\begin{equation}
\int \frac{dp_3}{2 \pi} \frac{e^{-i p_3 x_3}}{D_F(p)} \, = 
  \frac{e^{-2\hat{\omega}(p) x_3 }}{a \omega(p) \bar{\omega}(p)}
\end{equation}
where
\be
\omega^2(p) = \overline{m}^2(p) + \sum_{\nu=0}^2 \bar p_\nu ^2\; ,\quad 
\bar{\omega}^2(p) = \Big[\overline{m}(p) + \frac{2}{a}\Big]^2 + \sum_{\nu=0}^2 \bar p_\nu ^2 \; ,
\ee
\be
\!\!\!\!\overline{m}(p) = m_0 + \frac{a}{2} \sum_0^2 \hat p_n^2\; , \quad\;
a \hat{\omega}(p) =\frac{1}{2} \ln \Big[\frac{\bar\omega(p) + \omega(p)}{\bar\omega(p) - \omega(p)} \Big]\; . 
\ee
By using the above formulas, one finds that
\begin{equation}
  \begin{split}
    C_{{\cal O}}(x_3) = -\frac{3}{a^2 L_0} \sum_{n_0}\int dp_1\,dp_2 \; \bar C_{{\cal O}}(p) \; e^{-4\, \hat{\omega}(p) x_3 }
\end{split}
\end{equation}
where
\be\displaystyle
\bar C_{S}(p) = - \frac{4 \sum_{\nu=0}^2 \bar p_\nu ^2}{\omega^2(p) \bar{\omega}^2(p)}\,, \qquad
\bar C_{V_\mu}(p) = (1 - \delta_{\mu 3})\Big[\frac{1}{[\overline{m}(p) + 1/a]^2} - \frac{4\bar p_\mu^2}{\omega^2(p) \bar{\omega}^2(p)}\Big]\,,
\ee
\be
\hspace{-2.0cm} \bar C_{P}(p) = \frac{1}{[\overline{m}(p) + 1/a]^2}\, ,\;\;\; \qquad\!\!\!\!\! 
\bar C_{A_\mu}(p)\!= - \frac{\delta_{\mu 3}}{[\overline{m}(p) + 1/a]^2} + \frac{4 \sum_{\nu\neq\mu,3}\, \bar p_\nu ^2 }{\omega^2(p) \bar{\omega}^2(p)} \; .
\ee
Notice that $\bar C_S + \bar C_P = \bar C_{V_\mu} - \bar C_{A_\mu}$. For the shift vector $\bsxi=(1,0,0)$, the minimum of $\hat{\omega}$ is attained for
$(p_0,p_1,p_2) = (\frac{\pi}{2 L_0}, \frac{\pi}{2 L_0},0)$ for all correlators we are interested in. The tree-level values of the
screening masses are therefore all the same. They are given by the expression
\be
m^{\rm free}_{\cal O} = 4\, \hat{\omega} \Big(\frac{\pi}{2 L_0}, \frac{\pi}{2 L_0},0\Big)\; , 
\ee
whose values normalized to $2\pi T$ are listed, for practical convenience, in Table~\ref{tab:TreeM}
for the temporal extensions $L_0/a$ relevant to this paper.

\vspace{-0.25cm}

\bibliographystyle{JHEP}
\bibliography{bibfile.bib}

\end{document}